\documentclass[twocolumn,twocolappendix]{aastex631}

\newcommand{\kms}{km\,s$^{-1}$}
\usepackage{CJKutf8}

\shorttitle{Distant Echoes of GSE}
\shortauthors{Chandra et al.}

\graphicspath{{./}{fig/}}

\newcommand\nparent{200\,000}

\begin{document}

\defcitealias{Naidu2021}{N21}

\title{Distant Echoes of the Milky Way's Last Major Merger}

\author[0000-0002-0572-8012]{Vedant Chandra}
\affiliation{Center for Astrophysics $\mid$ Harvard \& Smithsonian, 60 Garden St, Cambridge, MA 02138, USA}

\author[0000-0003-3997-5705]{Rohan P. Naidu}
\altaffiliation{NASA Hubble Fellow}
\affiliation{MIT Kavli Institute for Astrophysics and Space Research, 77 Massachusetts Ave., Cambridge, MA 02139, USA}
\affiliation{Center for Astrophysics $\mid$ Harvard \& Smithsonian, 60 Garden St, Cambridge, MA 02138, USA}

\author[0000-0002-1590-8551]{Charlie Conroy}
\affiliation{Center for Astrophysics $\mid$ Harvard \& Smithsonian, 60 Garden St, Cambridge, MA 02138, USA}

\author[0000-0002-4863-8842]{Alexander P. Ji}
\affiliation{Department of Astronomy \& Astrophysics, University of Chicago, 5640 S Ellis Avenue, Chicago, IL 60637, USA}
\affiliation{Kavli Institute for Cosmological Physics, University of Chicago, Chicago, IL 60637, USA}

\author[0000-0003-4996-9069]{Hans-Walter Rix}
\affiliation{Max-Planck-Institut f{\"u}r Astronomie, K{\"o}nigstuhl 17, D-69117 Heidelberg, Germany}

\author[0000-0002-7846-9787]{Ana~Bonaca}
\affiliation{The Observatories of the Carnegie Institution for Science, 813 Santa Barbara Street, Pasadena, CA 91101, USA}

\author[0000-0002-1617-8917]{Phillip~A.~Cargile}
\affiliation{Center for Astrophysics $\mid$ Harvard \& Smithsonian, 60 Garden St, Cambridge, MA 02138, USA}

\author[0000-0002-6800-5778]{Jiwon~Jesse~Han}
\affiliation{Center for Astrophysics $\mid$ Harvard \& Smithsonian, 60 Garden St, Cambridge, MA 02138, USA}

\author[0000-0002-9280-7594]{Benjamin~D.~Johnson}
\affiliation{Center for Astrophysics $\mid$ Harvard \& Smithsonian, 60 Garden St, Cambridge, MA 02138, USA}

\author[0000-0001-5082-9536]{Yuan-Sen~Ting \begin{CJK*}{UTF8}{gbsn}(丁源森)\end{CJK*}}
\affiliation{Research School of Astronomy \& Astrophysics, Australian National University, Cotter Road, Weston, ACT 2611, Australia}
\affiliation{School of Computing, Australian National University, Acton ACT 2601, Australia}

\author[0000-0002-0721-6715]{Turner~Woody}
\affiliation{Center for Astrophysics $\mid$ Harvard \& Smithsonian, 60 Garden St, Cambridge, MA 02138, USA}

\author[0000-0002-5177-727X]{Dennis~Zaritsky}
\affiliation{Steward Observatory and Department of Astronomy, University of Arizona, Tucson, AZ 85721, USA}

\correspondingauthor{Vedant Chandra}
\email{vedant.chandra@cfa.harvard.edu}

\begin{abstract}

\noindent The majority of the Milky Way's stellar halo consists of debris from our Galaxy's last major merger, the \textit{Gaia}-Sausage-Enceladus (GSE). 
In the past few years, stars from GSE have been kinematically and chemically studied in the inner $30$~kpc of our Galaxy. 
However, simulations predict that accreted debris could lie at greater distances, forming substructures in the outer halo. 
Here we derive metallicities and distances using \textit{Gaia} DR3 XP spectra for an all-sky sample of luminous red giant stars, and map the outer halo with kinematics and metallicities out to $100$~kpc.
We obtain follow-up spectra of stars in two strong overdensities --- including the previously identified Outer Virgo Overdensity --- and find them to be relatively metal-rich and on predominantly retrograde orbits, matching predictions from simulations of the GSE merger. 
We argue that these are apocentric shells of GSE debris, forming $60-90$~kpc counterparts to the $15-20$~kpc shells that are known to dominate the inner stellar halo. 
Extending our search across the sky with literature radial velocities, we find evidence for a coherent stream of retrograde stars encircling the Milky Way from $50-100$~kpc, in the same plane as the Sagittarius Stream but moving in the opposite direction. 
These are the first discoveries of distant and structured imprints from the GSE merger, cementing the picture of an inclined and retrograde collision that built up our Galaxy's stellar halo. 
        
\end{abstract}

\keywords{Milky Way stellar halo (1060), Stellar streams (2166), Galaxy mergers (608), Milky Way formation (1053), Milky Way Galaxy (1054)}

\section{Introduction} \label{sec:intro}

The majority of our Galaxy's stellar halo was likely deposited by one massive merger event, the so-called \textit{Gaia}-Sausage-Enceladus (GSE, e.g., \citealt{Deason2013,Belokurov2018a,Helmi2018}). GSE was discovered as a distinct sequence in the color-magnitude diagram (CMD) of halo stars and as a coherent kinematic population in memory-preserving integrals of motion. 
Subsequent work has developed a picture of a merger $\approx 8-10$~Gyr ago that built up most of the present-day stellar halo \citep[e.g.,][]{Haywood2018, Gallart2019, Mackereth2019, Gallart2019, Lancaster2019, Bignone2019, Fattahi2019, Vincenzo2019, Naidu2020, Das2020, Bonaca2020, Naidu2021, Belokurov2022b}. To date, GSE has been extensively studied in the inner halo of the Milky Way, out to $\approx 30$~kpc. In this regime, the GSE debris is thoroughly phase-mixed, and must be identified via kinematic or chemical signatures. Consequently, models are relatively unconstrained regarding the precise orientation and dynamics of the merger. 

The hallmark of a major merger like GSE is rapid radialization, depositing stars on radial orbits with low azimuthal angular momentum and high eccentricity (e.g., \citealt{Amorisco2017}, \citealt{Naidu2021}, \citealt{Vasiliev2022}). The metallicity distribution function (MDF) of GSE sharply peaks at $\text{[Fe/H]} \approx -1.2$ and is well-fit by a simple chemical evolution model \citep[e.g.,][]{Hasselquist2020, Naidu2020, Feuillet2020, Horta2022a, Johnson2022, Limberg2022}. \cite{Naidu2020} report a population of retrograde stars dubbed `Arjuna' that closely matches this MDF, and argued that it may represent the early-stripped tail of GSE stars. The existence of this population hints at a retrograde orientation for the merger. \citet{Belokurov2022b} instead argue for a prograde configuration based on a mild prograde tilt of nearby ($d \lesssim 15$~kpc) GSE debris in the \textit{Gaia} DR3 RVS sample \citep{GaiaCollaboration2022,Katz2022}. Linking the orientation of present-day inner debris to the original trajectory of the merger is further complicated by any precession of the Milky Way disk since $z \approx 2$, which would change the net angular momentum vector of the Galaxy \citep[e.g.,][]{Dillamore2022,Dodge2022}

\citet[][hereafter \citetalias{Naidu2021}]{Naidu2021} ran a comprehensive suite of N-body simulations of the GSE merger, tailor-made to reproduce the kinematics of GSE and Arjuna stars from the H3 Survey \citep{Conroy2019b}. Their fiducial model consists of a $M_\ast = 5 \times 10^8~M_\odot$ merger beginning at $z \approx 2$, which started out with a tilted and retrograde orientation before rapidly radializing. No spatial constraints were used to orient their simulations --- the initially retrograde orientation is required to reproduce Arjuna's kinematics in the H3 Survey \citep{Naidu2020}. Nonetheless, their model reproduces the spatial orientation of the observed stellar halo out to $\approx 40$~kpc \citep{Juric2008,Xue2015,Das2016,Iorio2018,Iorio2019}, including a primary axis tilted off the plane of the Galactic disk \citep{Han2022b} and aligned with the so-called Hercules-Aquila Cloud (HAC; \citealt{Belokurov2007c,Simion2018,Simion2019}) and inner Virgo Overdensity (VOD; \citealt{Juric2008,Bonaca2012,Donlon2019}). {\cite{Li2016b} previously linked the orbit of the so-called Eridanus-Phoenix overdensity to the HAC and VOD as well, suggesting a common origin for all three structures.}

The HAC and VOD define a preferred spatial axis for GSE debris between $\approx 10-20$~kpc. One of the key predictions from \citetalias{Naidu2021} is a stellar stream of retrograde GSE stars beyond these distances, possessing a similar on-sky track to the Sagittarius stream, albeit moving in the opposite direction. This debris should also contain observable apocentric overdensities, with stars `echoing' back and forth between regions of pile-up on the sky. The prediction of debris beyond $\approx 50$~kpc is generic to simulations of this merger, with more radially anisotropic configurations producing larger fractions of distant debris \citep[e.g.,][]{Bignone2019,Elias2020}.

At more than four times the distance from the GSE-associated VOD lies the \textit{outer} Virgo Overdensity (OVO; \citealt{Sesar2017a}), a hitherto mysterious structure discovered via RR Lyrae stars (RRL) at $\approx 80$~kpc. It is interesting to consider that this outer overdensity lies in one of two preferred octants for debris from the GSE merger, perhaps linking it to the extended outer streams predicted by simulations. {Likewise, \cite{Carlin2012} used orbital information to associate the VOD to the southern Pisces Overdensity at $\approx 70$~kpc, another possible distant remnant of GSE \citep{Sesar2007,Kollmeier2009,Sesar2010a}. } 

However, disentangling and interpreting these structures has thus far been challenging in the absence of kinematic information. Although RRL have long been standard outer halo tracers due to their precise distances, at $80$~kpc they have typical \textit{Gaia} DR3 tangential (proper motion) velocity uncertainties $\sigma_\mathrm{v} \approx 300$~\kms{}.
It is also challenging to measure radial velocities for RRL since they exhibit pulsation phase-dependent RV variations that need to be modelled. This makes them unsuitable to search for cold velocity substructures in the outer halo, motivating the need for more luminous tracers like red giant stars. 

Red giant branch (RGB) stars are luminous tracers with characteristic tangential velocity uncertainties $\sigma_\mathrm{v} \lesssim 30$~\kms{} at $80$~kpc, an order of magnitude improvement over RRL. This comes at the cost of distance accuracy, and their isochrone distances particularly depend on the assumed metallicity. The third data release from \textit{Gaia} has revolutionized this latter effort by releasing low-resolution ($R \approx 50-100$) `XP' prism spectra for $\gtrsim 200$~million stars brighter than $G \lesssim 17.6$ \citep{Montegriffo2022a,DeAngeli2022}. These spectra enable us to measure metallicities for red giant stars, deriving much more accurate distances than previously possible. This dataset enables us to, for the first time, comprehensively map the outer halo with 5D and --- for the subset with follow-up and literature radial velocities --- 6D kinematic information. 

In this work, we use luminous red giants with 5D and 6D kinematics to discover echoes of GSE in the outer halo of the Galaxy. We use \textit{Gaia} DR3 astrometry and spectrophotometry to construct a pure sample of red giants out to $100$~kpc. We augment this sample with radial velocities from follow-up spectra and existing spectroscopic surveys, creating an unprecedented dataset with 6D kinematics in the outer halo. We describe our sample construction, metallicity measurements, and follow-up spectroscopy in $\S$\ref{sec:data}. We present our analysis of apocentric overdensities in $\S$\ref{sec:analysis:ovo}, and extend our search for GSE debris across the sky in $\S$\ref{sec:analysis:allsky}. We conclude by discussing our results and future lines of inquiry in $\S$\ref{sec:discussion}. 

\section{Data and Analysis}\label{sec:data}

\subsection{Selecting RGB Stars}

We begin by assembling a sample of RGB stars using \textit{Gaia}~DR3 astrometry and unWISE infrared photometry \citep{Mainzer2014,Schlafly2019,GaiaCollaboration2021,Lindegren2021,GaiaCollaboration2022}. We query \textit{Gaia}~DR3 for sources with $\varpi < 0.4$~mas, $\mu < 5$~mas/yr, and $\mid b \mid > 20^\circ$ to remove obvious nearby dwarfs and mask out the MW disk. \textit{Gaia} parallaxes for distant giants are typically low-significance measurements that cannot reliably be turned into distances --- most have $\varpi / \sigma_\varpi \lesssim 1$. However, these parallaxes are still powerful to remove contamination from foreground dwarf stars. For a star with a given apparent $\text{G}$ magnitude and $\text{BP}-\text{RP}$ color, we predict the parallax of a $\log{g} = 4$ dwarf using MIST isochrones \citep{Choi2016}. We divide the difference between the predicted and observed parallax by the predicted parallax uncertainty for the star's $\text{G}$ magnitude, deriving a significance statistic $\chi_{\mathrm{\varpi}}$ that encodes how many standard deviations lie between the observed parallax and the prediction for a dwarf. Using stars from the H3 Spectroscopic Survey \citep{Conroy2019b} as a guide, we find that removing stars with $\chi_{\mathrm{\varpi}} < 2$ results in a $\approx 90\%$ pure sample of stars with $\log{g} < 3.5$. 

We further purify this sample by applying a broad color cut in the $\left(\text{BP}-\text{RP},\text{RP}-\text{W1}\right)$ color space to remove dwarfs based on their \textit{WISE} infrared colors \citep{Conroy2018,Conroy2021}, which increases the sample purity to $\gtrsim 95\%$. Our resulting parent sample consists of $\approx \nparent{}$ RGB stars brighter than $G \leq 17.65$ in the color range $1.3 \leq \text{BP}-\text{RP} \leq 3$, i.e., K and M type giants. 

\subsection{5D Giants: Metallicities and Distances with \textit{Gaia}~XP}\label{sec:data:xpmet}

\begin{figure}
    \centering
    \includegraphics[width=\columnwidth]{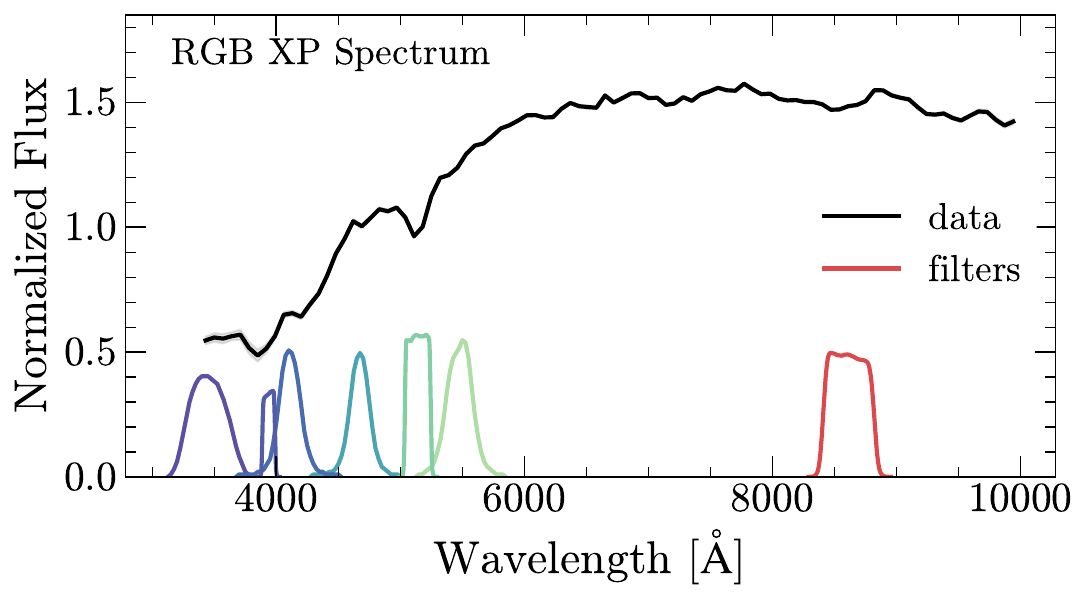}
    \includegraphics[width=\columnwidth]{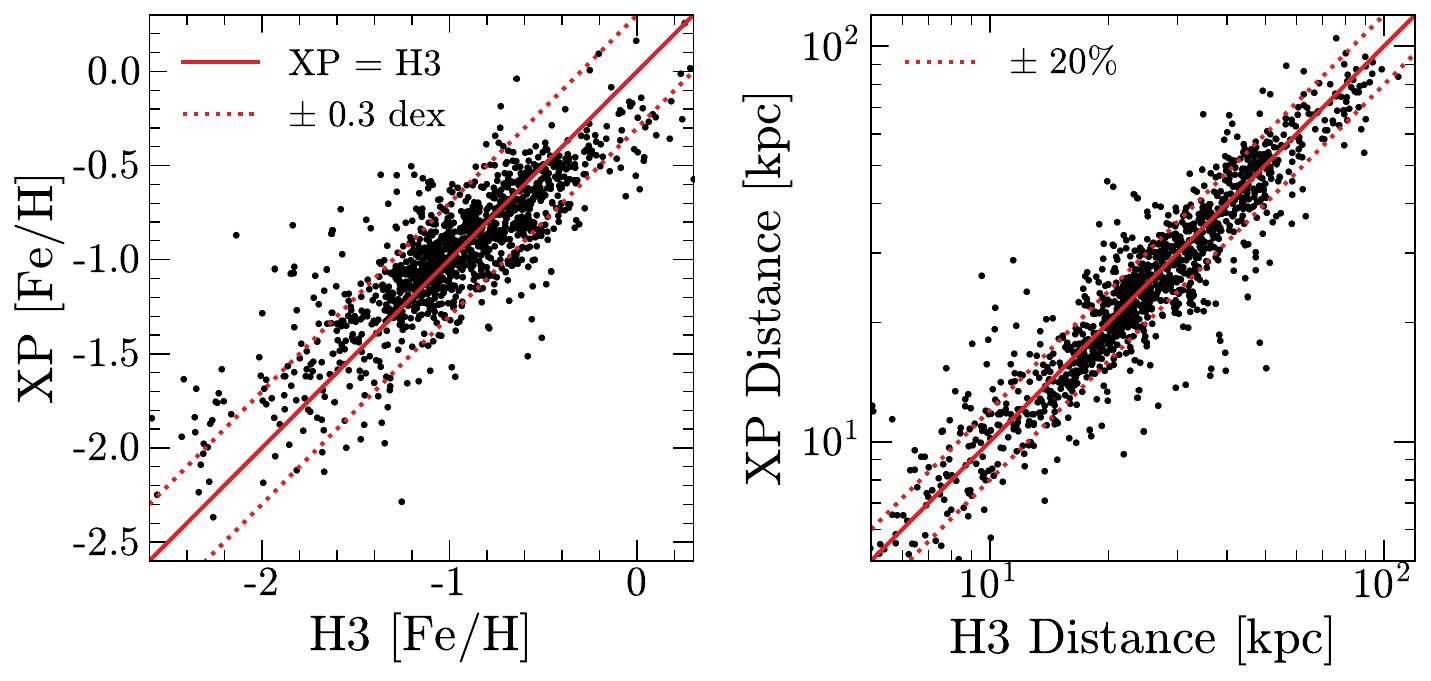}
    \caption{Top: Example of an observed \textit{Gaia} DR3 XP spectrum of an RGB star. We display the medium-band photometric filters --- Str\"omgren with round peaks, JPLUS with flat peaks --- for which we synthesize photometry from these spectra, in order to predict their metallicities and distances. Bottom: Validating our XP-derived metallicities and distances by comparing against spectroscopic measurements from the H3 Survey. We use two-fold cross-validation here, i.e., we train the model on half the data and evaluate it on the other half, and vice versa.}
    \label{fig:xp_validation}
\end{figure}

Mapping the spatial and orbital properties of giants requires an estimate of their distance, and photometric isochrone distances are heavily dependent on the assumed metallicity \citep[e.g.,][]{Conroy2021}. We therefore derive spectro-photometric metallicities for our giants using low-resolution `XP' prism spectra from \textit{Gaia}~DR3 \citep{DeAngeli2022,Montegriffo2022a}. In order to compress the information contained in these continuously-represented spectra into astrophysically interpretable quantities, we compute integrated photometry in several narrow band filters using the \texttt{GaiaXPy} utility \citep[v1.0.2,][]{Mieres2022,Montegriffo2022b}. These filters are drawn from the J-PLUS and Str{\"o}mgren systems, and are centred on absorption features that are sensitive to stellar parameters like metallicity \citep[Figure~\ref{fig:xp_validation}, see e.g.,][]{Stromgren1966, Bailer-Jones2004,Marin-Franch2012}. As our feature set, we use colors of the following bands relative to the synthesized $g$ band: J0395, J0515, J0861, u, v, b, y. We de-redden these colors using the dust maps of \cite{Schlegel1998} --- re-normalized by \cite{Schlafly2011} --- and restrict our sample to $E(B-V) \leq 0.3$. In practice this removes a small fraction of our sample, since most high-extinction regions are already masked out via the $\mid b \mid > 20^\circ$ cut on the parent sample. 

To map this multidimensional feature set to metallicities, we use high-resolution spectroscopic data from the H3 Survey \citep{Conroy2019b}, which have metallicities estimated using the \texttt{MINESweeper} routine \citep{Cargile2020}. Ideally, we would build a generative model of the spectral energy distributions using theoretical spectra and estimate metallicities in a full fitting framework. However, this method is sensitive to offsets between the theoretical and observed colors, which are challenging to isolate and correct. Instead, we opt for a discriminative regression in which we directly map colors to metallicities in observed space. This circumvents systematic offsets in the theoretical models or filter curves, at the cost of a well-defined error model for the predicted metallicities. This trade-off is quite justified for our present science case \citep[see][for an application of similar techniques]{Rix2022}. 

We select $\approx 1500$ stars from our parent sample that were also observed by the H3 Survey, and consequently have high-resolution [Fe/H] measurements. These stars span a wide range of parameters $3800 \lesssim T_\mathrm{eff}/\mathrm{K} \lesssim 5000$, $0 \lesssim \log{g} \lesssim 4$, and $-3.0 \lesssim \mathrm{[Fe/H]} \lesssim 0.3$. We train a two-layer neural network with 16 neurons in each layer to map our XP-synthesized color space to H3-measured [Fe/H] and $\log{g}$ \citep{Pedregosa2011}. We experimented with less flexible models like linear and quadratic regression, and found that the neural network produced a noticeably better root-mean-square prediction error on test data that was withheld from the training set. We use two-fold cross-validation --- training on 1/2 of our dataset and predicted on the remaining 1/2, and vice-versa --- to increase our confidence that the chosen regression model is not over-fitting the training data. Our final model predicts H3 [Fe/H] to within 0.3~dex on average, and $\log{g}$ to within 0.1~dex, with negligible bias (Figure~\ref{fig:xp_validation}, bottom left). Due to the purity of our parent giant sample the vast majority of the data have XP-inferred $\log{g} < 3.5$, but we use the $\log{g}$ predicted by our model to further clean the sample and remove non-giants with XP-predicted $\log{g} > 3.5$ {($\approx 1\%$ of the stars)}. 

We estimate $G$-band distance moduli --- and consequently distances --- to all $\approx \nparent{}$ giants in our sample using a 10~Gyr MIST isochrone spline-interpolated to each star's XP-inferred [Fe/H] and $\log{g}$. This assumption of a fixed age is well-founded \citep[e.g.,][]{Bonaca2020}, and a factor of two change in the assumed age affects the resulting distances by less than $10\%$. A comparison to spectroscopic distances from the H3 Survey demonstrates that we obtain unbiased distances with $\lesssim 20\%$ scatter out to 100~kpc (Figure~\ref{fig:xp_validation}, bottom right). 

\newpage

\subsection{6D Giants}\label{sec:data:rvs}

\subsubsection{Literature Radial Velocities}

In order to derive 6D phase-space information for a subset of our sample, we cross-match our giants to catalogs from the Sloan Extension for Galactic Understanding and Exploration (SEGUE; \citealt{Lee2008,Yanny2009,Eisenstein2011,Alam2015}), and Large Sky Area Multi-Object Fibre Spectroscopic Telescope (LAMOST; \citealt{Cui2012,Zhao2012,Xiang2019}), as well as the H3 Survey \citep{Conroy2019b,Cargile2020}. 
We use a version of the SEGUE catalog with parameters estimated via the \texttt{MINESweeper} routine \citep{Cargile2020}, which should have more robust [$\alpha$/Fe] values than the default pipeline, and includes spectrophotometric distances.
We do not utilize the \textit{Gaia} DR3 Radial Velocity Survey (RVS; \citealt{Katz2022}) here, since it contains near-zero stars beyond $50$~kpc due to its shallow depth. 
This yields $\approx 12\,000$ giants with 6D kinematic information and spectroscopic metallicities, {$\approx 1100$ of which are beyond 40~kpc}. We refer to both the `5D giant' and `6D giant' datasets in this paper.

\subsubsection{Follow-up Spectroscopy with MIKE}\label{sec:data:mike}

To more comprehensively sample the two outer halo overdensities studied in this work, we obtained follow-up spectroscopy for $29$ XP-selected giants using the Magellan Inamori Kyocera Echelle (MIKE; \citealt{Bernstein2003}) spectrograph on the 6.5\,m Magellan Clay telescope at Los Campanas Observatory in Chile (PIs Naidu and Ji). We used $\approx 10$~minute exposures with the 0\farcs7 slit to attain signal-to-noise (S/N) $\gtrsim 10$ on the red end of the $R \approx 20\,000$ spectra. We reduced the data using the CarPy utility, and derived radial velocities using Payne4MIKE with typical uncertainties $\approx 0.5$~\kms{} \citep[][Ji et al., in preparation]{Kelson2003,Ting2019}. Further details on the target selection and analysis are described in $\S$~\ref{sec:analysis:ovo}.

Our MIKE spectra have SNR~$\approx 10$ in the calcium triplet (CaT) region, but only SNR~$\approx 1$ in the bluer ($\approx 5000$\,\AA) portions of the spectrum, hampering a sophisticated fit to the entire optical spectrum. 
We therefore derive metallicities for our MIKE stars using the CaT calibration of \cite{Carrera2013}. 
We fit Voigt profiles to each of the three CaT absorption lines using \texttt{lmfit} \citep{Newberg2009} and derive their equivalent widths. 
The \cite{Carrera2013} calibration relies on the summed equivalent widths and absolute V-band luminosity ($M_{\text{V}}$) of the star. 
We make a first estimate of $M_{\text{V}}$ using the XP-inferred metallicity and MIST isochrones. 
We then compute the spectroscopic metallicity following \cite{Carrera2013}, and re-compute $M_{\text{V}}$.
This process is iterated until the metallicity and $M_{\text{V}}$ converge, which in practice occurs within 5 iterations. 

Our MIKE CaT metallicities are broadly consistent with the XP metallicities for these stars, with a median absolute deviation of $\approx 0.2$~{dex} and a bias of $\approx -0.1$~{dex}, and with the MIKE metallicities being slightly more metal-poor than XP (recall that the XP metallicities were trained on H3 Survey data, $\S$\ref{sec:data:xpmet}). 
This offset is unsurprising given that our adopted CaT metallicity calibration does not take into account $\alpha$-enhancement, whereas H3 jointly fits for [Fe/H] and [$\alpha$/Fe] with high-resolution spectra of the Mgb triplet region \citep{Cargile2020}. 
We utilize the MIKE metallicities (and their corresponding isochrone distances) for the remainder of this work except where otherwise noted, while cautioning that there is a slight offset in the MIKE metallicity scale compared to XP and the H3 Survey. 

\subsubsection{Orbital Parameters}

We compute orbital parameters for stars with 6D phase-space information using \texttt{gala} \citep{gala,adrian_price_whelan_2020_4159870}, adopting the default \texttt{MilkyWayPotential}. For the Solar position with respect to the Galaxy, we adopt a radius $r_\odot = 8.122$~kpc \citep{GravityCollaboration2018}, and a height above the mid-plane $z_\odot = 20.8$~pc \citep{Bennett2019}. We use a right-handed Galactocentric cartesian coordinate frame with a solar position $\mathbf{x}_\odot = (-8.12, 0.00, 0.02)$~kpc, and solar velocity $\mathbf{v}_\odot = (12.9, 245.6, 7.8)$~$\mathrm{km\,s^{-1}}$ \citep{Reid2004,Drimmel2018,GravityCollaboration2018}. 
% Actions are computed using the ``St\"ackel Fudge'' \citep{Binney2012, Sanders2012} with \texttt{galpy} \citep{Bovy2015}. 
Orbital parameters are computed after numerically integrating the orbits with a timestep of 1~Myr over 2.5 Gyr.

\section{Apocenter Pileup: Echoes in the Galactic North and South}\label{sec:analysis:ovo}

In this section, we use our \textit{Gaia} DR3 sample to identify RGB stars in the Outer Virgo Overdensity and Pisces Overdensity, and characterize their kinematics and chemistry with follow-up MIKE spectroscopy. We argue that these overdensities represent apocentric shells from the GSE merger, matching several key predictions from simulations of the merger. Finally, we use these stars --- along with the \citetalias{Naidu2021} simulations --- to measure the metallicity gradient of the GSE progenitor out to the edge of its disk. 

\subsection{Spectroscopic Follow-up of the Outer Virgo Overdensity and its Southern Counterpart}\label{sec:analysis:ovo:specfollow}

\begin{figure*}
    \centering
    \includegraphics[width=0.7\textwidth]{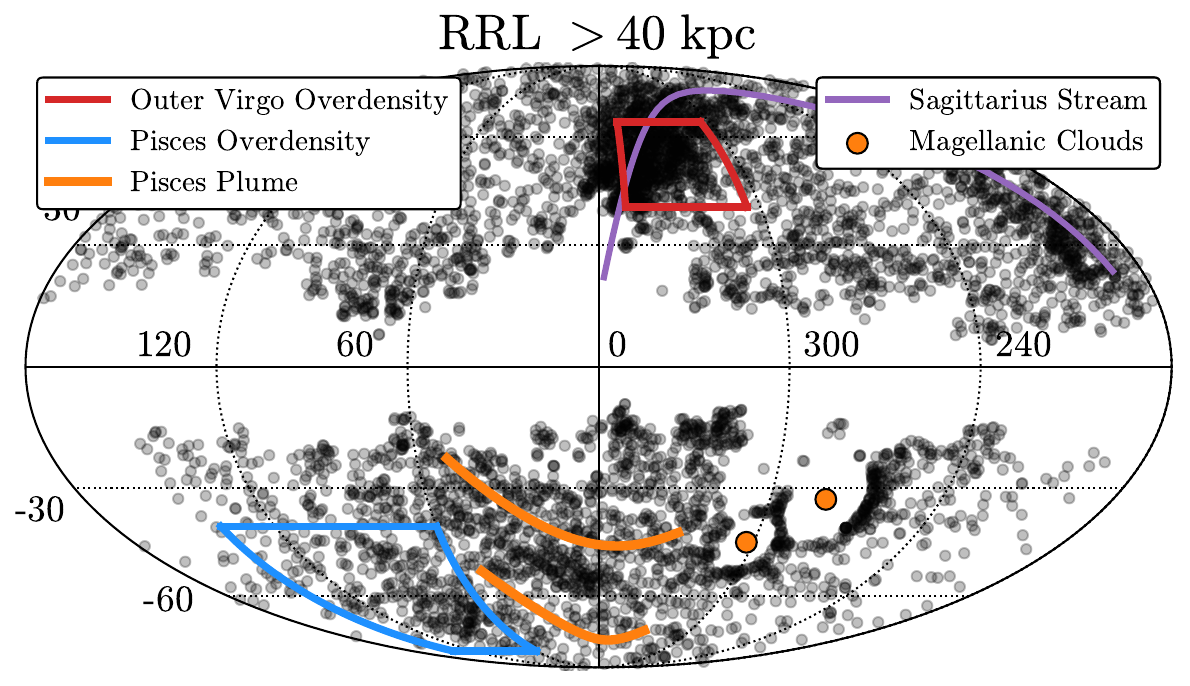}
    \includegraphics[width=0.7\textwidth]{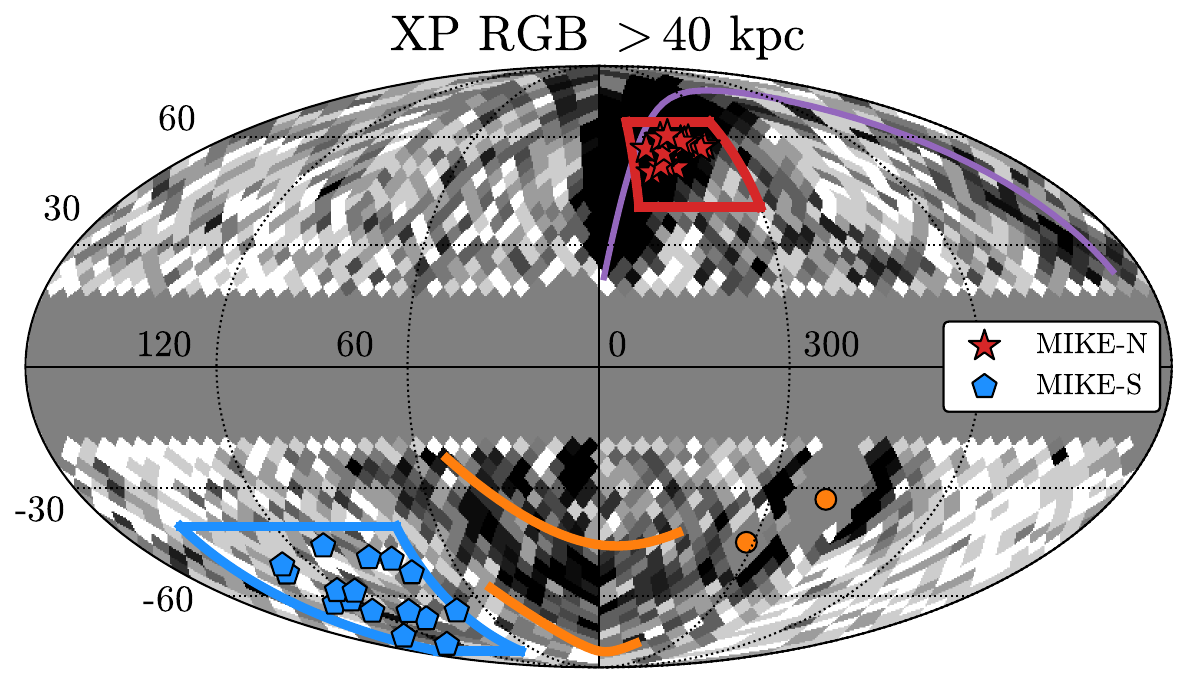}
    \caption{The spatial distribution of outer halo substructure as seen by RRL and RGB stars. Top: All-sky distribution of \textit{Gaia} DR3 RRL beyond 40~kpc, with the overdensities studied in this work highlighted by red and blue boxes. 
    Bottom: Density distribution of our all-sky \textit{Gaia} XP sample of RGB stars beyond 40~kpc. We overlay our spectroscopic targets in the MIKE-North (Outer Virgo Overdensity) and MIKE-South (Pisces Overdensity) regions. There are observed overdensities of RRL and RGB stars in both these locations, albeit weaker and more extended in the south. }
    \label{fig:allsky}
\end{figure*}

{Figure~\ref{fig:allsky} illustrates the $d > 40$~kpc outer halo with $\approx 6600$ RRL from \textit{Gaia} DR3 \citep[top panel,][]{Clementini2022}, and $\approx 8500$ of our 5D XP RGB stars (bottom panel)}. 
Several key structures and overdensities are highlighted. 
The Outer Virgo Overdensity (OVO) was first identified by \cite{Sesar2017a} using RRL from Pan-STARRS \citep{Sesar2017}. 
It was described as a strong excess of Pan-STARRS RRL clustered around $(\alpha,\delta) \approx (207,-7)$ at a distance of $\approx 80$~kpc, with a line-of-sight Gaussian extent of $\sim 4$~kpc \citep{Chambers2016,Sesar2017,Sesar2017a}. 
The northern map is dominated by distant arm of Sagittarius \citep{Majewski2003}, whereas in the south the elongated Pisces Plume is prominent \citep{Belokurov2019a,Conroy2021}. 
{We overlay a polynomial track $\pm 10^\circ$ around $l = -60 + 0.1 b + 0.007 b^2$ to guide the eye along the Plume. The origin of the Pisces Plume is uncertain, particularly whether it contains debris stripped from the Magellanic Clouds, or if it mainly represents the dynamical friction wake of the Large Magellanic Cloud \citep{Belokurov2019a,Conroy2021}}.

Apart from the Plume, there is another prominent southern overdensity that we highlight in blue. 
This structure is spatially coincident with the previously identified Pisces Overdensity \citep[][]{Sesar2007,Watkins2009,Kollmeier2009,Nie2015}, and like the OVO lies in one of the preferred octants for GSE debris. 
{We delineate this structure with $60^\circ < l < 140^\circ$ and $-80^\circ < b < -40^\circ$, enclosing the previously identified regions of the overdensity.
Our selection is significantly wider than past surveys of the PO, since the overdensity in our K~giant sample spans a larger swath on the sky.} 

\begin{figure}
    \centering
    \includegraphics[width=\columnwidth]{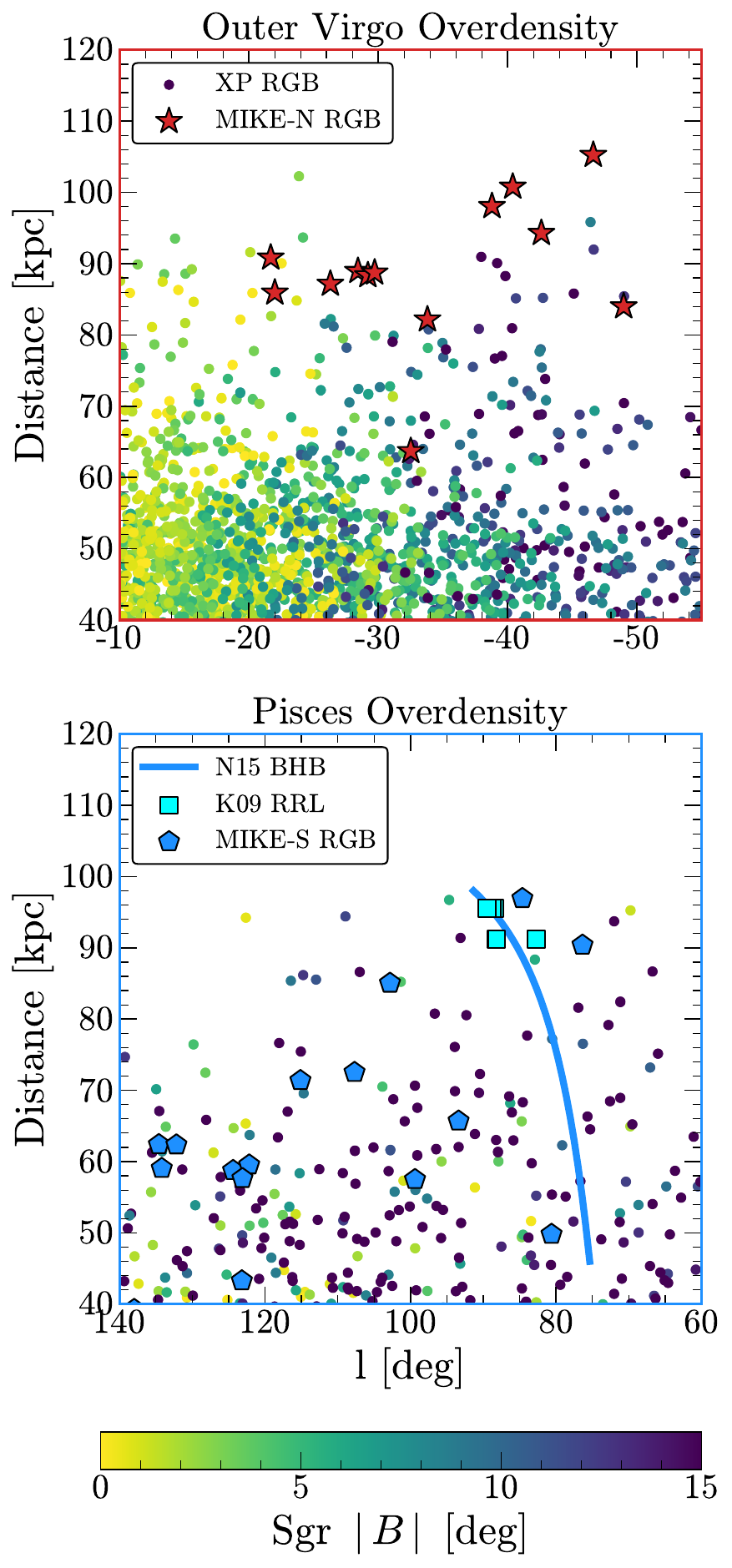}
        \caption{
        The Galactic longitude -- distance distributions of XP RGB stars beyond 40~kpc in the Outer Virgo Overdensity (top panel) and Pisces Overdensity (bottom panel) selection boxes shown on the all-sky maps of Figure~\ref{fig:allsky}. The OVO appears in the top panel as a distinct cloud of stars at $\approx 80$~kpc, offset by $\approx 10^\circ$ from the Sagittarius plane. The Pisces Overdensity RRL reported by \citet{Kollmeier2009} are overlaid in the bottom panel, along with the BHB distance gradient measured by \citet{Nie2015}. In both panels we overlay our MIKE targets using their spectroscopic distances.
        }
        \label{fig:ovo_rrl}
\end{figure}

The OVO and Pisces Overdensity are shown in Galactic longitude--distance coordinates in Figure~\ref{fig:ovo_rrl}. 
We color the stars by their angular separation from the Sagittarius stream track \citep{Law2010a}. 
The OVO stands out as a clear overdensity off the plane of Sagittarius, and at almost twice the distance of Sagittarius debris in the plane of the sky.
The corresponding southern overdensity is weaker and more extended.
However, previously-identified RRL are insufficient to kinematically disentangle and identify populations due to their faintness and consequently large tangential velocity uncertainties. 
We therefore set out to obtain follow-up spectroscopy of RGB stars in both overdensities to obtain 6D kinematics and more precise metallicities for a representative sample of stars. 

In the north, we selected 13 stars from our 5D XP giants dataset that lie within $\approx 5^\circ$ of the OVO with XP-inferred distances of $\approx 70-100$~kpc, bracketing the OVO RRL. 
We applied no selections on XP-inferred metallicity and kinematics for this northern sample. 
We performed a corresponding selection for 16 stars in the south, guided by the overdensity of stars visible on the all-sky RRL and RGB maps (Figures~\ref{fig:allsky} \& \ref{fig:ovo_rrl}). 
Since the southern overdensity is weaker and more extended, we applied selection in Solar reflex-corrected proper motions of $\mu_{\alpha}^\ast > 0$~mas\,yr$^{-1}$ ($\approx 50$\% of the sample) to increase the likelihood of selecting retrograde stars.
This was guided by the fact that there were two giants in our 6D RGB star catalog (with literature radial velocities) lying in the Outer Virgo Overdensity, and both had retrograde kinematics. 
We therefore targeted stars with similar kinematics in the south, to ascertain if there was an associated population there. 
We account for this selection bias in the subsequent interpretations. 

Our spectroscopic targets are shown in Galactic coordinates in Figure~\ref{fig:allsky}, and are also overlaid on the distance maps of Figure~\ref{fig:ovo_rrl}.  
The bottom panel of Figure~\ref{fig:ovo_rrl} demonstrates that although our southern targets are in a similar region of sky as the previously-identified Pisces Overdensity, they are closer and more extended in Galactic longitude. 
We henceforth refer to the respective MIKE samples as MIKE-N (northern OVO) and MIKE-S (southern counterpart around the Pisces Overdensity) to avoid ambiguity. 
MIKE-N is an unbiased survey of stars lying in the Outer Virgo Overdensity, whereas MIKE-S is a PM-targeted search for stars near the Pisces Overdensity that have a higher likelihood of being retrograde. 

\subsection{Kinematics of the Outer Overdensities}\label{sec:analysis:kin}

Following the spectroscopic reduction and fitting described in $\S$\ref{sec:data:mike}, we are armed with 6D kinematics and chemistry for our MIKE targets. 
The MIKE-N stars have median MIKE (XP) [Fe/H]~$\approx -1.45$ ($-1.31$), whereas the MIKE-S stars have median MIKE (XP) [Fe/H]~$\approx -1.36$ ($-1.28$).
This relatively high metallicity already points to a massive progenitor galaxy for these stars --- in the outer halo, the two most likely sources of metal-rich stars are GSE or Sagittarius. 
{We note that \cite{Watkins2009} found a slightly lower [Fe/H]~$\approx -1.5$ in the Pisces Overdensity (MIKE-S) region, although their sample was more spatially concentrated and centered around $85$~kpc compared to our broader sample around $60$~kpc (see the bottom panel of Figure~\ref{fig:ovo_rrl}).}

{\cite{Conroy2019a} find that stars in the H3 Survey beyond $30$~kpc have a richly structured metallicity distribution, with both metal-rich ([Fe/H]~$\approx -1.2$) and metal-poor ([Fe/H]~$\approx -2.0$) components. 
Subsequent work has argued that the majority of the [Fe/H]~$\approx -1.2$ stars can be attributed to the GSE merger \citep{Naidu2020,Naidu2021}. 
Therefore, the metallicity of these overdensities suggests that they might share a common origin with the bulk of the field halo stars at these distances. 
GSE stars are expected to be tidally stripped over a range of radii, plausibly populating both the field halo and the apocentric overdensities presented in this work. 
}

\begin{figure}
    \centering
    \includegraphics[width=\columnwidth]{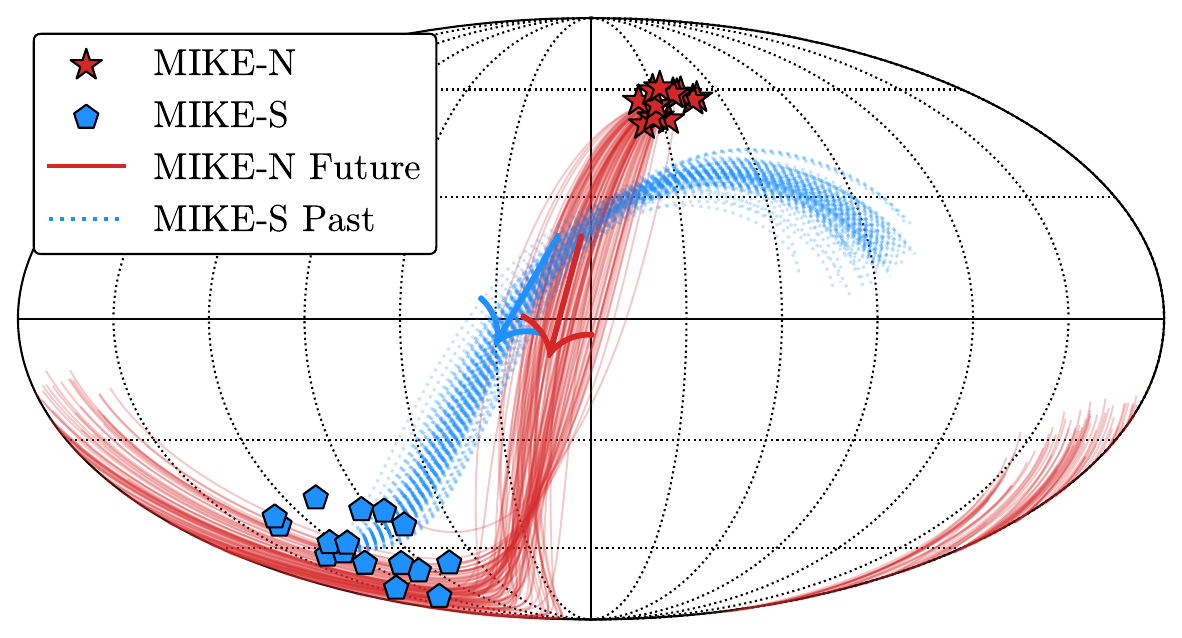}
    \includegraphics[width=\columnwidth]{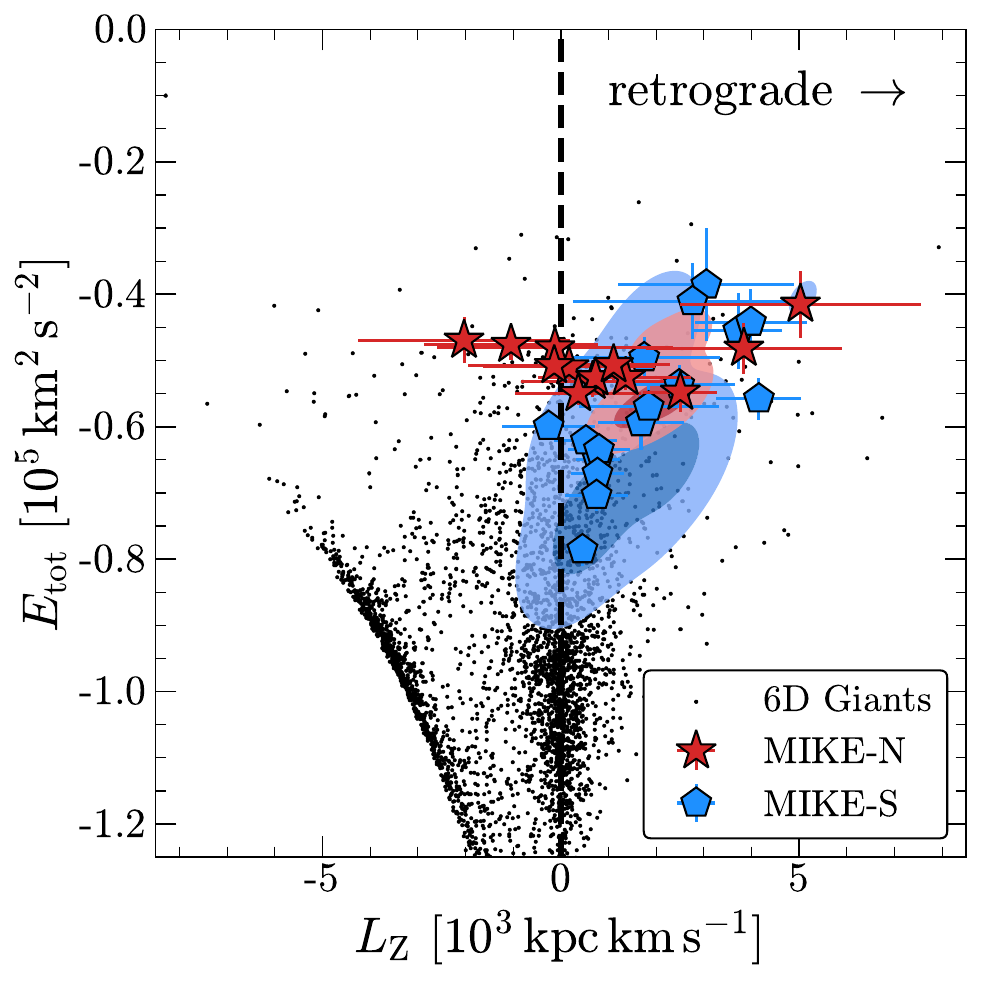}
    \caption{Top: Future (past) orbits for the MIKE-N (MIKE-S) stars targeted in the Outer Virgo Overdensity (Pisces Overdensity), integrated over 1 Gyr with arrows denoting the general direction of these orbits on-sky. Bottom: MIKE targets in $L_\mathrm{Z}$-$E_{\mathrm{tot}}$ space, with the \textit{Gaia} XP 6D giant sample shown for comparison (black dots). The density distribution for \citetalias{Naidu2021} GSE simulation particles selected from analogous sky regions are shown using contours with corresponding colors. The MIKE stars in both hemispheres are overwhelmingly retrograde. Note that that the MIKE-S stars had a proper motion selection criterion that boosts the likelihood of being retrograde, whereas the MIKE-N stars were blindly targeted within the Outer Virgo Overdensity ($\S$\ref{sec:analysis:ovo:specfollow}).}
    \label{fig:mike_iom}
\end{figure}

Figure~\ref{fig:mike_iom} illustrates the kinematics of our MIKE sample. The inverse-variance weighted mean 6D phase-space coordinates of MIKE-N (MIKE-S) stars are integrated forward (backward) in time for 1 Gyr in the \texttt{gala} \texttt{MilkyWayPotential} with 1 Myr timesteps. 
The motivation for these respective timescales is that in the \citetalias{Naidu2021} simulations, the northern debris temporally precedes the southern debris as the GSE progenitor falls into the Milky Way. 
We sample the mean phase space coordinates from their uncertainty distributions (assuming Gaussian uncertainties on the observed coordinates and kinematics) and display 100 realizations of these orbits in the top panel of Figure~\ref{fig:mike_iom}.
The MIKE-N and MIKE-S populations --- and by extension the underlying Outer Virgo Overdensity and at least the retrograde portion of the Pisces Overdensity --- appear kinematically associated, moving in a common orbital plane. 
The correspondence is not exact, suggesting that these populations may have been stripped at different times during the merger, and hence occupy slightly different orbital planes and phases. 
Furthermore, the assumed potential is static and likely inaccurate in detail \citep[e.g.,][]{Erkal2019a,Koposov2022}. \
Regardless, these orbits suggest that the northern stars will occupy a region near the southern overdensity in $\approx$ a Gyr, and conversely that the  southern stars used to reside near the northern overdensity $\approx$ a Gyr ago. 
These are possible signatures of a pair of apocentric pileups, evoking an image of stars `echoing' between these overdensities over time. 

The bottom panel of Figure~\ref{fig:mike_iom} shows the azimuthal angular momentum versus total orbital energy of these stars. 
Stars targeted in the northern Outer Virgo Overdensity are overwhelmingly retrograde, despite being selected without any kinematic bias. 
The two stars with slightly prograde kinematics could be interlopers from the nearby Sagittarius stream, as they also have modestly negative $L_{\text{Y}}$ (see the top panel of Figure~\ref{fig:ovo_rrl}). 
The southern stars were kinematically selected to have a high likelihood of being retrograde on the basis of proper motions, and are indeed all so. 
The retrograde nature of these stars strongly argues against an association with the Sagittarius stream, which is highly prograde. 
These stars are also inconsistent with the extremely negative $L_{\text{Y}}$ angular momentum of Sagittarius, having a median $L_{\text{Y}} \approx 1 \times 10^3\,\mathrm{kpc\,km\,s}^{-1}$. 

For comparison, we select star particles from the fiducial \citetalias{Naidu2021} GSE merger simulations lying in analogous northern and southern regions beyond $\approx 40$~kpc, and overlay their distribution with contours in Figure~\ref{fig:mike_iom}. 
There is a reasonable correspondence between the models and data. 
In particular, the models predict that the southern stars lie at lower energies (smaller apocenters) than the northern stars, which is reflected in our data. 
Our MIKE-N targets have a median distance $\approx 80$~kpc, whereas the MIKE-S stars lie $\approx 60$~kpc away. 

We estimate how close these stars are to their apocenters with the statistic $f_{\text{apo}} = (r_{\text{gal}} - r_{\text{peri}}) / (r_{\text{apo}} - r_{\text{peri}})$. 
The stars have a median $f_{\text{apo}} \approx 0.94$ in the North and $f_{\text{apo}} \approx 0.90$ in the South. 
This affirms our conclusion that these stars are piling up near their apocenters. 
This is analogous to our current understanding of the \textit{inner} Virgo Overdensity and Hercules Aquila Cloud, which are likewise thought to consist of GSE stars near apocenter \citep{Simion2019,Iorio2019,Naidu2021,Perottoni2022,Han2022b}. 

Previously, \cite{Zaritsky2020} used data from the H3 Survey to independently identify a kinematically cold structure of 15 stars near the Pisces Overdensity at $40 \lesssim d/\mathrm{kpc} \lesssim 80$. This structure lies near our MIKE-S sample, although the \cite{Zaritsky2020} stars are more clustered around the canonical Pisces Overdensity RRL than our more extended MIKE-S selection. They argued that these stars may be debris stripped from the Small Magellanic Cloud, based on their negative Galactocentric radial velocities and [Fe/H]~$\approx -1.4$ mean metallicity. Furthermore, the stars are linked in position and velocity with the gaseous Magellanic stream \citep{Nidever2008}. 

As noted by \cite{Zaritsky2020}, the distances and angular momenta of these stars are somewhat discrepant compared to the bulk of simulated SMC debris from \cite{Besla2013}, although more recent simulations may be in better agreement \citep{Lucchini2021a}. The structure reported by \cite{Zaritsky2020} is in fact modestly retrograde, and overlaps with the MIKE-S stars in orbital energy and between $L_{\mathrm{Z}} \approx 0-1.5 \times 10^3\,\mathrm{kpc\,km\,s}^{-1}$ (see Figure~\ref{fig:mike_iom}). We therefore speculate that the stars identified by \cite{Zaritsky2020} may also be associated with the distant GSE debris discovered in our work, with a correspondence on-sky, in kinematics, and in mean metallicity. However, this region of sky requires more comprehensive spectroscopy to disentangle the various structures --- chiefly the Pisces Overdensity and Pisces Plume (see Figure~\ref{fig:allsky}) --- and ascertain whether these stars are more plausibly associated with GSE or the Small Magellanic Cloud. 

\section{Distant Streams of GSE}\label{sec:analysis:allsky}

In $\S$\ref{sec:analysis:ovo} we presented follow-up spectroscopy of a pair of on-sky overdensities that correspond to the most prominent GSE debris piling up near apocenter. 
In this section, we extend the search for debris across the sky using our dataset of RGB stars with 6D kinematics via a cross-match to literature radial velocities. 
We identify a population of retrograde stars beyond $40$~kpc that seem to form a coherent track on the sky, aligned with the apocentric overdensities described in $\S$\ref{sec:analysis:ovo}. 
We suggest that these stars are likewise structured remnants from the merger, matching expectations of GSE based on their kinematics and chemistry. 

\subsection{Selection with 6D Giants}

\begin{figure}
    \centering
    \includegraphics[width=\columnwidth]{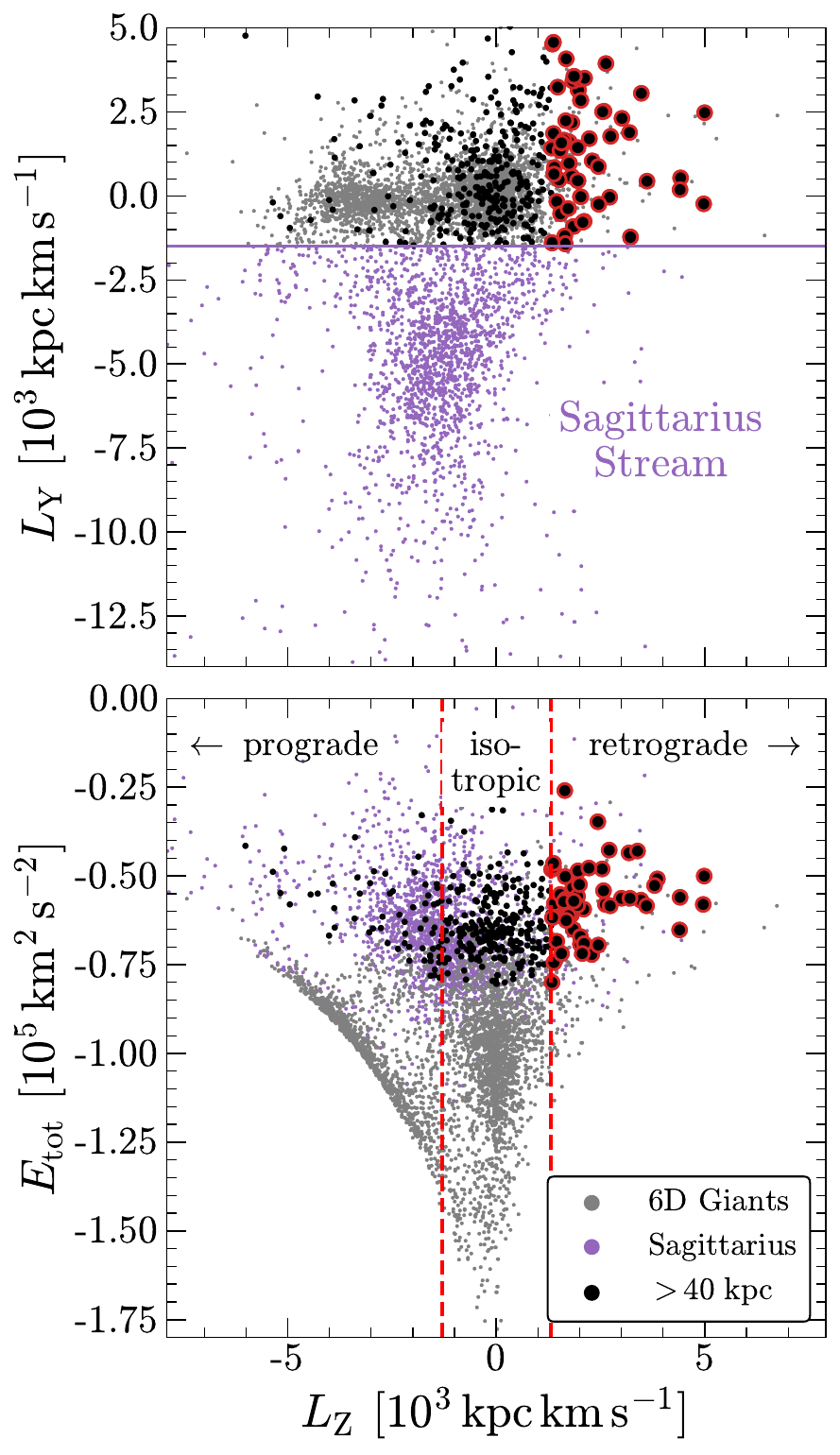}
    \caption{Isolating distant and retrograde debris in our 6D giant dataset with literature RVs. Top: While members of the distant GSE debris and Sagittarius have considerable overlap in their $L_\mathrm{Z}$--$E_{\mathrm{tot}}$ distribution, they are quite distinct in the orientation of their orbital plane, reflected in $L_\mathrm{Y}$. This allows us to separate the Sagittarius stream with with a simple $L_\mathrm{Y}$ cut, following \citet{Johnson2020a}. Bottom: The 6D giants in energy-angular momentum space, highlighting the distant $d > 40$~kpc sub-sample and separating stars into isotropic and retrograde bins, which are subsequently shown on-sky in Figure~\ref{fig:lz_maps}.}
    \label{fig:lz_selection}
\end{figure}

\begin{figure*}
    \centering
    \includegraphics[height=0.3\textwidth]{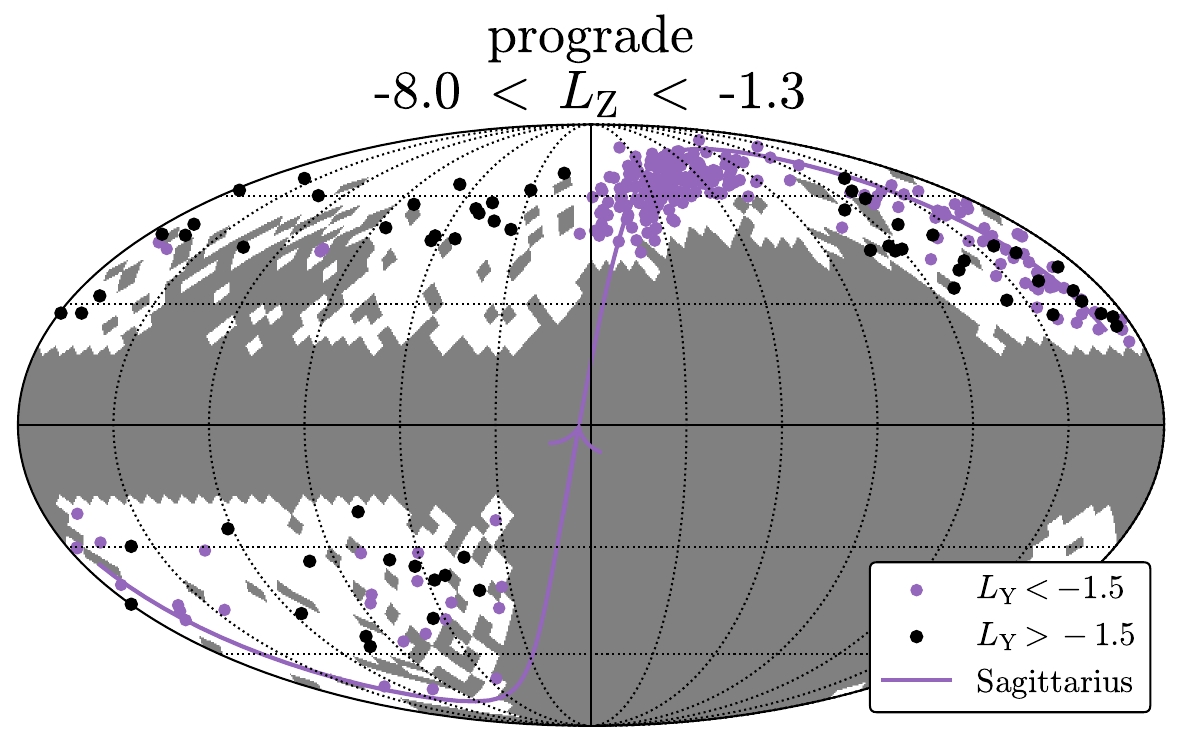}
    \includegraphics[height=0.3\textwidth]{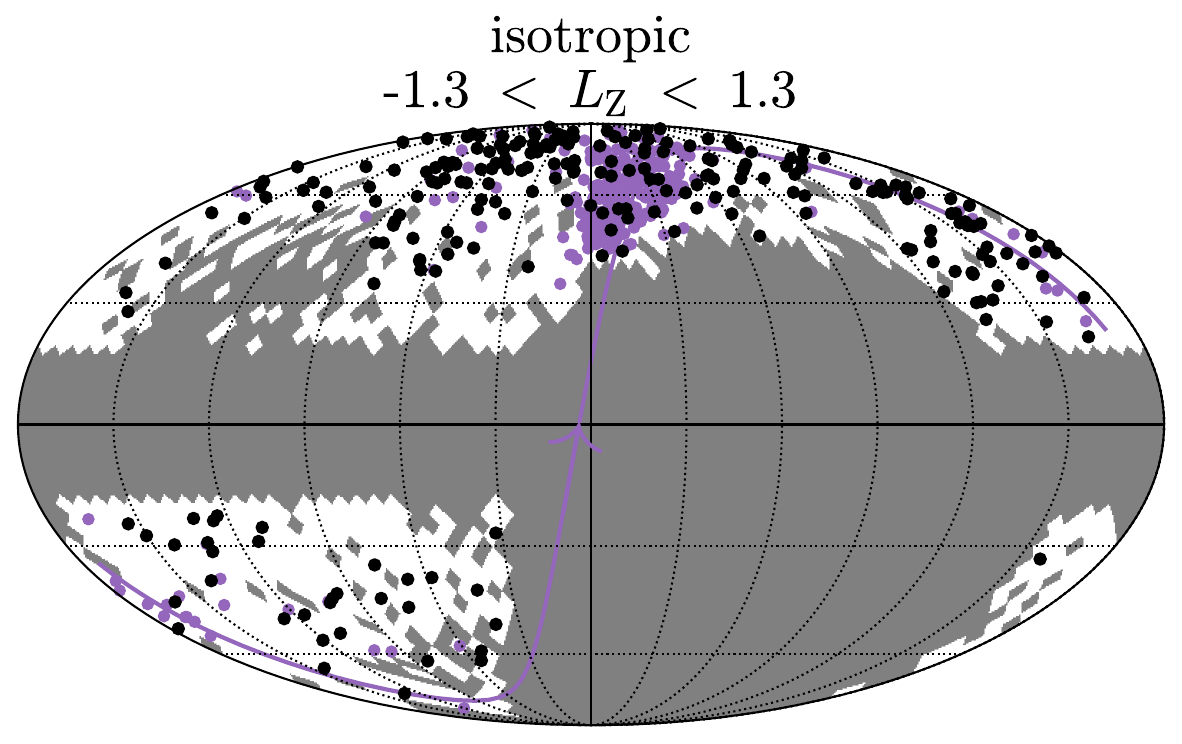}
    \includegraphics[height=0.3\textwidth]{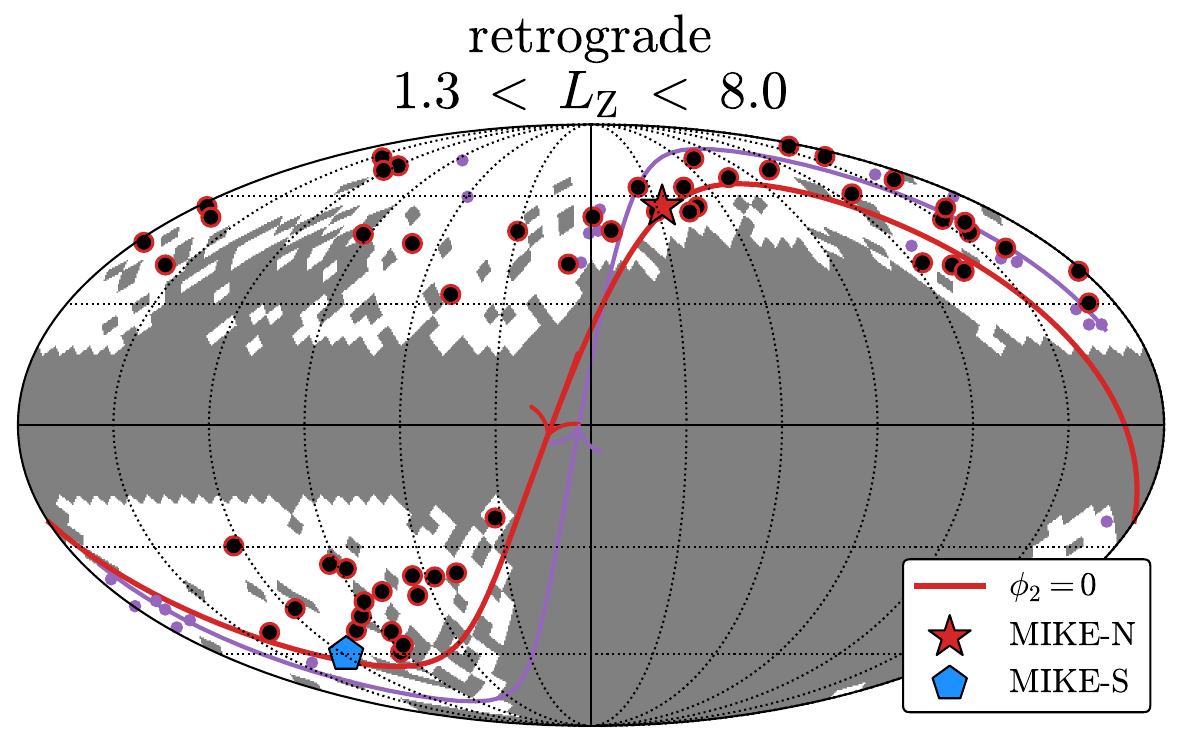}
    \includegraphics[height=0.3\textwidth]{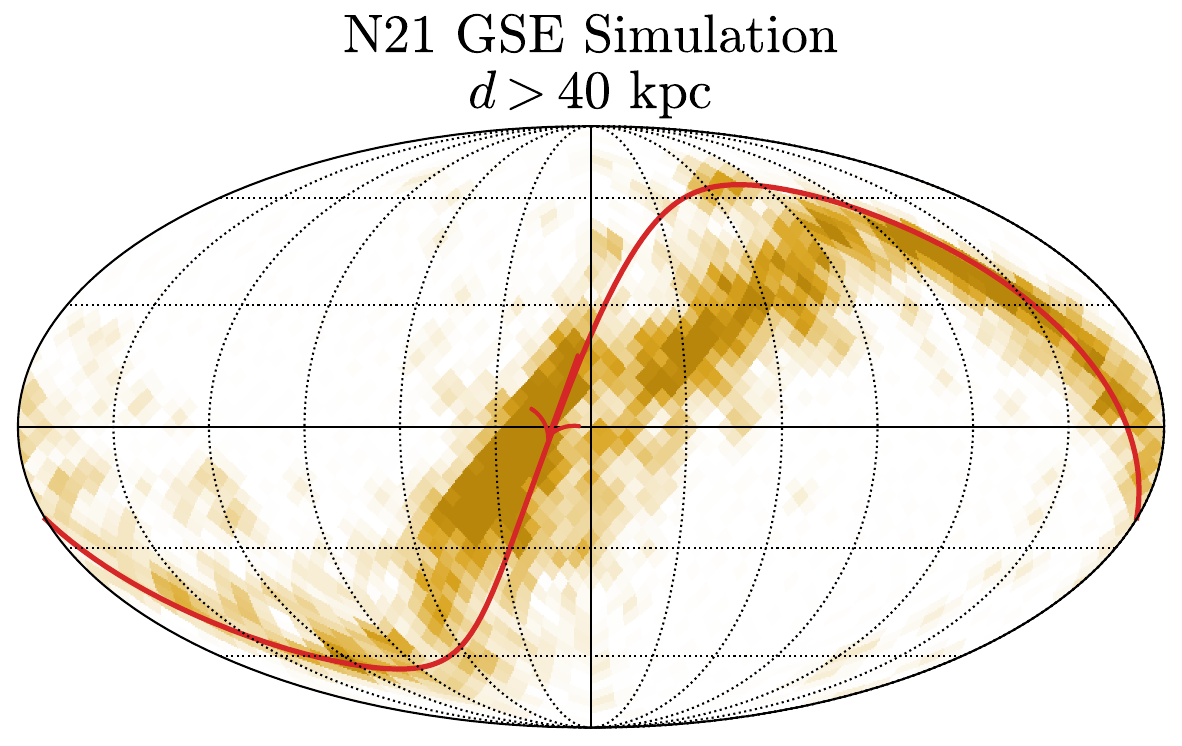}
    \caption{On-sky distribution of distant 6D giant stars, split into three subsets of azimuthal angular momentum $L_{\mathrm{Z}}$ in units of $10^3\,\text{kpc\,km\,s}^{-1}$: stars on prograde orbits, stars on isotropic orbits, and stars with retrograde orbits (see Figure~\ref{fig:lz_selection}). The selection function of the underlying spectroscopic surveys, and our Galactic latitude cut to mask out the MW disk, is shaded in grey. Sagittarius stream stars selected via $L_\mathrm{Y}$ are shown in purple. Whereas the stars with prograde and isotropic orbits are uniformly distributed across the survey footprints --- apart from Sagittarius, which dominates these panels --- the retrograde stars show a suggestive track on the sky, to which we fit the great circle shown in red. This track coincides with the MIKE-N (Outer Virgo Overdensity) and MIKE-S (Pisces Overdensity) stars, whose median positions are overlaid. A red (purple) arrow indicates the on-sky orbital motion of the retrograde (Sagittarius) stars, emphasizing that the spatially overlapping debris moves in opposite directions. In the bottom right panel we show GSE-only particles from the merger simulation of \citetalias{Naidu2021}, selected to lie beyond 40~kpc.}
    \label{fig:lz_maps}
\end{figure*}

We begin with the 6D giant sample in the energy-angular momentum space (bottom panel of Figure~\ref{fig:lz_selection}). 
Three structures dominate this space: the prograde and rotationally supported disk, the eccentric and radial GSE locus around $L_{\text{Z}} \approx 0$, and the cloud of Sagittarius stream stars at higher energies. 
We disentangle and excise the Sagittarius stars using an $L_{\text{Y}}$ cut following \cite{Johnson2020a}, coloring them separately in subsequent plots (Figure~\ref{fig:lz_selection}, top). 
Finally, we select stars with XP-inferred distances $d > 40$~kpc.

The resulting sample of distant stars is dominated by Sagittarius, with the remaining stars spanning a swatch of azimuthal angular momentum. 
We split these stars into prograde, isotropic, and retrograde $L_{\text{Z}}$ bins and display their spatial distribution in Figure~\ref{fig:lz_maps}. 
The footprint of our underlying spectroscopic surveys is shown in white, whereas grey regions denote 13~$\text{deg}^2$ healpixels containing zero spectra in the parent spectroscopic catalogs. 
The non-Sagittarius prograde stars are scattered across the sky (top left panel), although even our conservative $L_{\text{Y}}$ cut appears to have missed some tightly clustered Sagittarius stars in the top-right portion of the map. 
The bulk of Sagittarius stars fall into our isotropic $L_{\text{Z}}$ bin (top right panel), with non-Sagittarius stars uniformly distributed across the sky.
Finally, the retrograde selection entirely excludes the Sagittarius stream based on kinematics (bottom left panel).
Compared to the isotropic stars, the retrograde stars suggest a stream-like structure in the distant halo, aligned with the MIKE-N and MIKE-S samples presented in $\S$\ref{sec:analysis:ovo}. 
These stars lie in a plane similar to that of the Sagittarius stream, but move in the opposite direction. 
The appearance of this structure is certainly confounded by the selection footprint of the underlying (northern) spectroscopic surveys, although even just the northern $b \gtrsim 20^\circ$ stars exhibit a spatial asymmetry in the retrograde selection compared to the isotropic one. 
There is likely dominant GSE debris in both these panels, since even beyond $50$~kpc about half of the GSE debris is predicted by \citetalias{Naidu2021} to be radial with low angular momentum. 
However, we focus here on the retrograde population because it can be clearly differentiated from the field stellar halo.

To quantify the orbital plane of this structure, we fit the distribution of retrograde stars with a great circle, minimizing the mean squared angular deviation of these stars from the great circle track. 
We use this fitted track to define a coordinate frame $(\phi_1, \phi_2)$ aligned with the great circle (red line in the bottom panels of Figure~\ref{fig:lz_maps}). 
The overlaid track visually emphasizes how this sample of kinematically-selected stars forms a coherent track on the sky, passing through the Outer Virgo Overdensity and Pisces Overdensity. 
This structure has likely escaped detection thus far due to its spatial overlap with the Sagittarius stream, which dominates these regions of the sky.
Of the 6D giants beyond $40$~kpc, $\approx 60\%$ belong to the Sagittarius stream selection in $L_{\text{Y}}$, whereas only $\approx 7\%$ belong to the retrograde structure identified here. 

We qualitatively compare our result with simulation-based predictions from \citetalias{Naidu2021} in the bottom right panel of Figure~\ref{fig:lz_maps}. 
We perform the same $d > 40$~kpc selection as the data and display the on-sky density of simulated star particles, overlaying the fitted track described above. 
We emphasize here that the \citetalias{Naidu2021} simulations did not utilize any spatial information to constrain their model, yet the predicted track of distant GSE debris is quite consistent with our data. 
This lends further credence to the proposed retrograde orientation of the merger, building on prior evidence that GSE produced the $d \approx 20$~kpc Hercules Aquila Cloud and (inner) Virgo Overdensity \citep{Simion2019,Perottoni2022}. 

\subsection{Stream Coordinates and Chemistry}\label{sec:analysis:allsky:chemistry}

\begin{figure*}
    \centering
    \includegraphics[width=\textwidth]{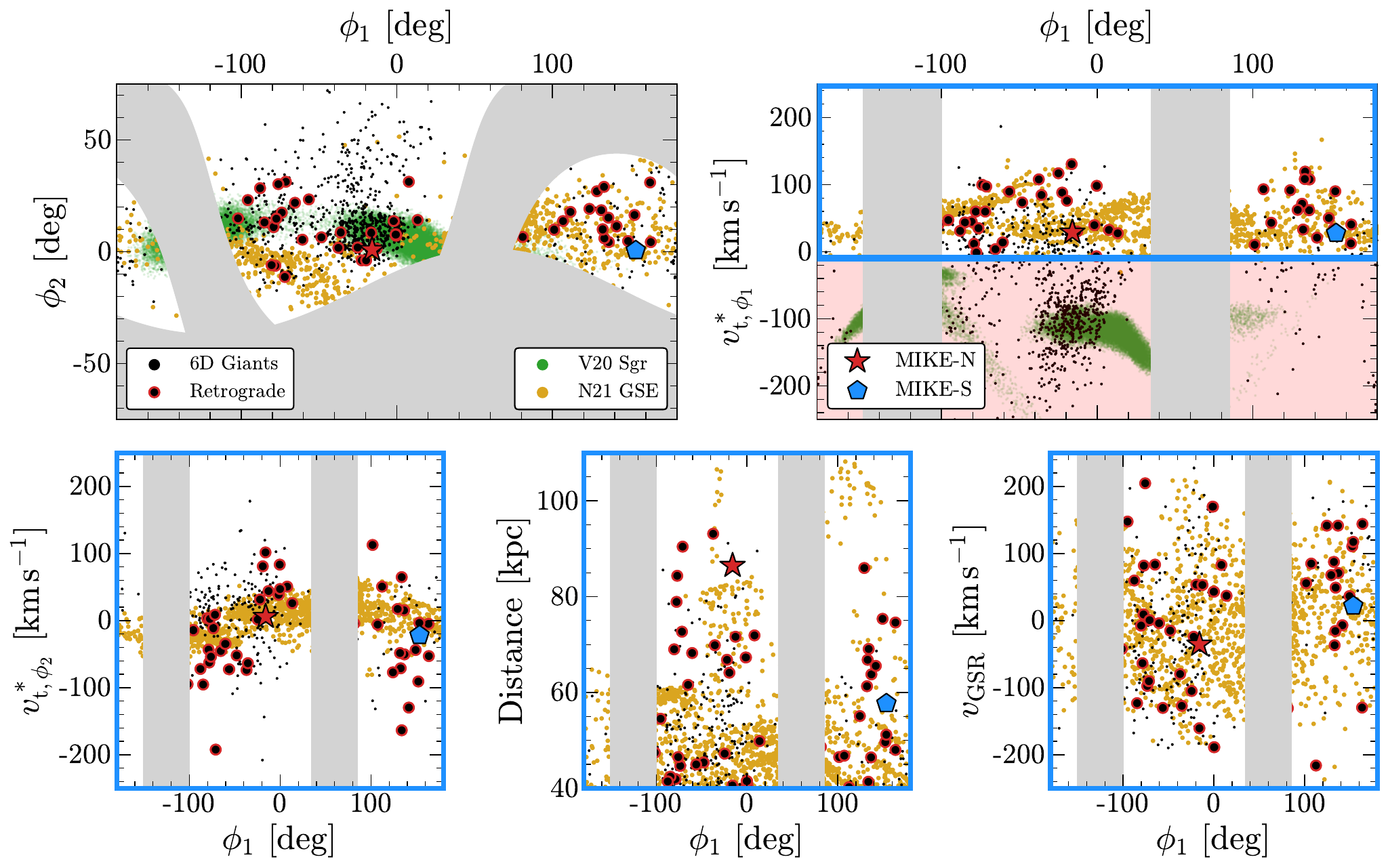}
    \caption{Comparison of the distant debris identified in this work to simulations of GSE and Sagittarius, shown in the transformed great circle coordinate frame. We show all distant 6D giants from our sample in black, and retrograde giants are outlined in red. GSE simulation particles from \citetalias{Naidu2021} are overlaid in gold, and Sagittarius simulation particles from \cite{Vasiliev2021a} in green. Grey regions mask out the main survey selection functions: the Galactic latitude cut of $\mid b \mid > 20^\circ$, and the southern declination limit of the underlying spectroscopic catalogs. A simple selection of $v_{\mathrm{t, \phi_1}} > 0$ isolates presumed members of GSE debris. In the bottom row of panels, we only show observed and simulated stars that satisfy this criterion, in the transverse (relative to the stream track) proper motion, distance, and line-of-sight velocity planes. For reference, the mean parameters of the apocentric overdensities followed up with MIKE are overlaid (see $\S$\ref{sec:analysis:ovo}).}
    \label{fig:phi1_summary}
\end{figure*}

Figure~\ref{fig:phi1_summary} illustrates our sample in the transformed great circle $(\phi_1,\phi_2)$ coordinate frame.
The entire distant 6D giant sample is shown without the Sagittarius $L_{\mathrm{Y}}$ cut applied, to establish the clear difference between GSE and Sagittarius in our data. 
We show simulation particles of the GSE merger from \citetalias{Naidu2021}, along with simulated Sagittarius stars from \citet[][]{Vasiliev2021a}. 
We select retrograde 6D giants that lie within $\phi_2 \leq 35^\circ$ of the great circle frame --- in practice, this spatial cut removes only $8$ field retrograde stars that lie elsewhere on the sky. 
We apply a further purity cut of $v_{\mathrm{t, \phi_1}} > 0$ to select stars that move along the great circle --- this removes another $3$ stars, leaving 49 stars in the distant retrograde sample.   
The stars that fail this cut are likely field halo stars that happen to be retrograde, or genuine stream stars with large proper motion errors.
{Their XP metallicity distribution is peaked at [Fe/H]~$\approx -1.2$, consistent with the field halo at these distances \citep{Conroy2019a}.}
In the bottom panels of Figure~\ref{fig:phi1_summary} we only show stars that satisfy this criterion.

The top right panel of Figure~\ref{fig:phi1_summary} shows tangential velocity along the $\phi_1$ coordinate, emphasizing how different the proper motions of the retrograde giants are from Sagittarius, despite lying in a similar on-sky plane.  
Although there is generally good agreement between our data and the \citetalias{Naidu2021} simulations in all these panels, the distribution of $\phi_2$ tangential velocities is hotter in the data than the simulations (bottom left panel). 
This could reflect some perturbative process not considered in the simulations, like the growing MW disk or the influence of Sagittarius and the LMC \citep[e.g.,][]{Law2010,Gomez2015,Erkal2016,Koposov2022}. 
The \citetalias{Naidu2021} simulations predict distant $d \gtrsim 50$~kpc debris at two location along this stream track, both of which we identify in the data as the Outer Virgo Overdensity (MIKE-N, $\phi_1 \sim 0^\circ$) and Pisces Overdensity (MIKE-S, $\phi_1 \sim 130^\circ$)  respectively (bottom middle panel). 
The disordered but net $\approx 0$~\kms{} Galactocentric radial velocities are typical of `shells' of merger debris in simulations \citep[bottom right panel, e.g.,][]{Hernquist1992,Pop2018,Dong-Paez2022}. 

\begin{figure}
    \centering
    \includegraphics[width=\columnwidth]{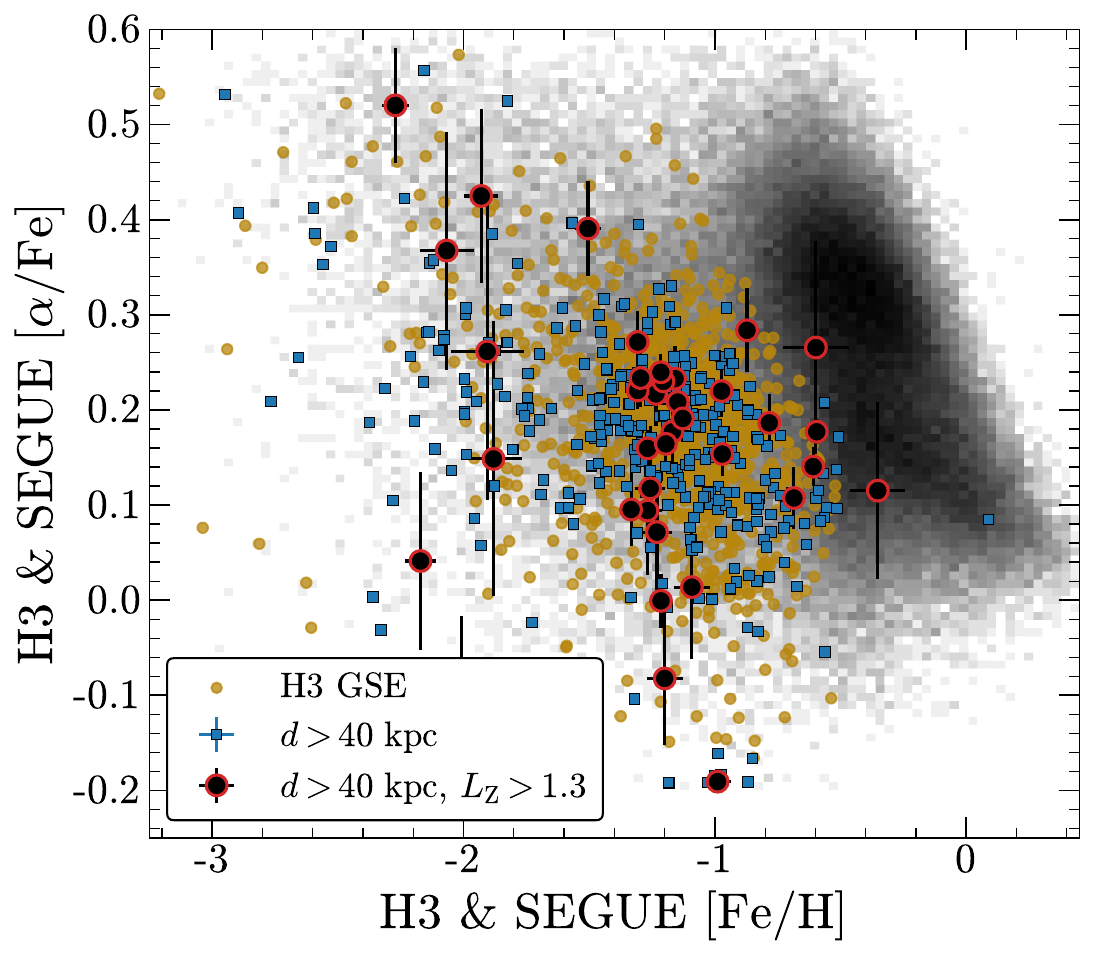}
    \caption{Distribution of the [Fe/H] \emph{vs.} [$\alpha$/Fe] abundances among the distant (blue squares) and retrograde (black and red circles) stars in our 6D giants dataset. We use stars with spectra from the H3 Survey and the SEGUE survey; both have been fitted using the techniques of \cite{Cargile2020} and lie on a similar abundance scale. For reference, we overlay in gold H3 GSE giants selected following \cite{Naidu2020}, with the log-scaled background histogram showing all stars in the H3 Survey.}
    \label{fig:retro6d_tw}
\end{figure}

We finally explore this all-sky retrograde debris in chemistry. 
We utilize the subset of retrograde 6D giants shown in Figure~\ref{fig:lz_maps} that have data from H3 and SEGUE. 
Both the H3 and SEGUE datasets were fitted using the \texttt{MINESweeper} routine to derive [Fe/H] and [$\alpha$/Fe], and therefore share the same abundance scale \citep[more details in ][]{Cargile2020}. 
We illustrate this chemical space in Figure~\ref{fig:retro6d_tw}. 
The background log-scaled histogram shows all stars in the H3 survey, and the golden points show GSE giants selected from H3 using the cuts of \cite{Naidu2020}. 
The retrograde giants we identify in this work match the GSE chemical sequence, clustering around its locus. 
We note that much of the outer halo shares a similar locus, perhaps in part because GSE itself forms much of the diffuse outer halo. 
There are a few retrograde stars off the GSE track with low [Fe/H] and low [$\alpha$/Fe], plausibly stars from past minor mergers with smaller dwarf galaxies \citep[e.g.,][]{Cohen2009,Cohen2010}.

\subsection{Metallicity Gradient of GSE's Outer Disk}

\begin{figure}
    \centering
    \includegraphics[width=\columnwidth]{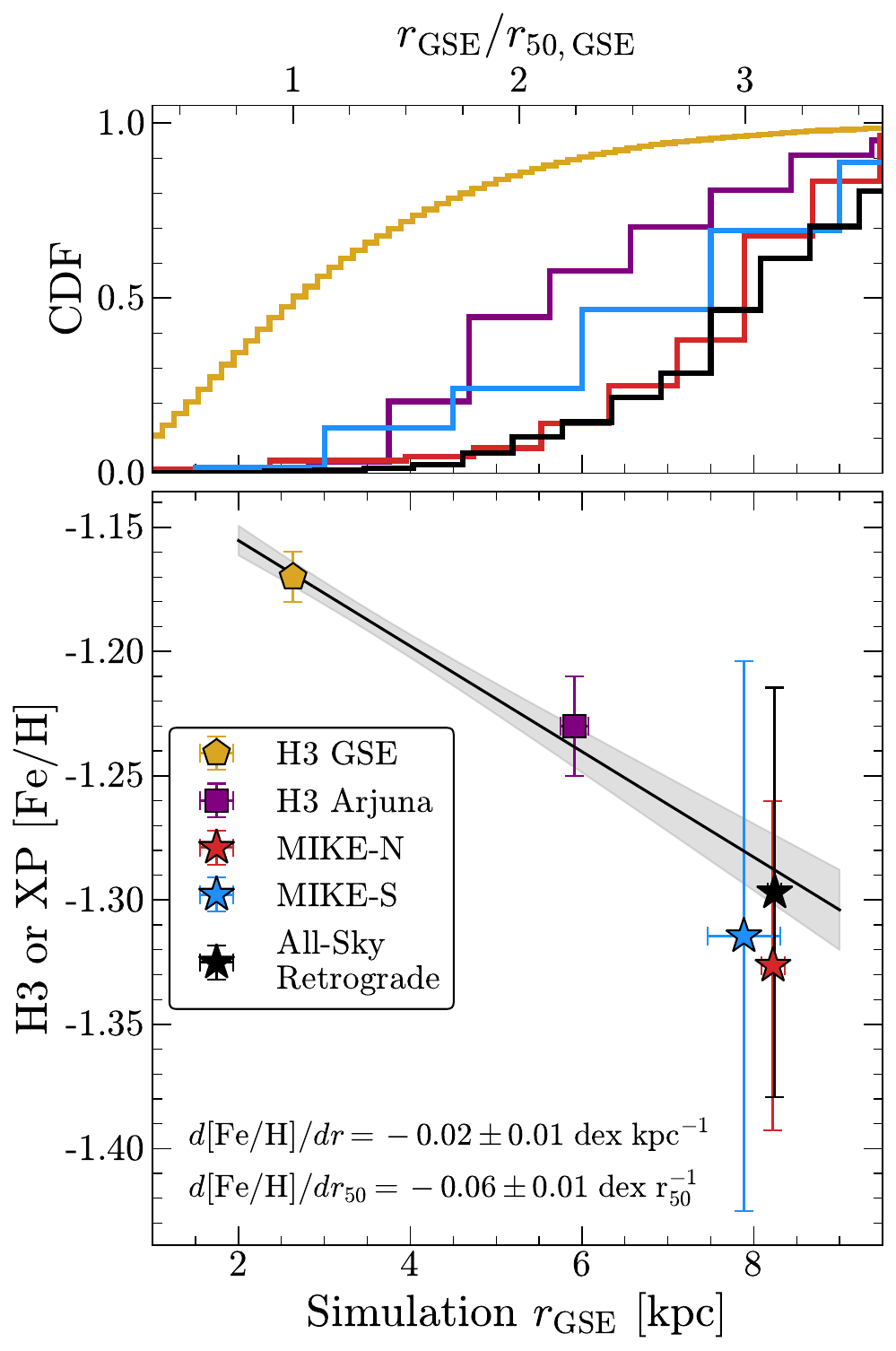}
    \caption{Measuring the metallicity gradient of the GSE progenitor to three times its half-mass radius. We use simulations of \citetalias{Naidu2021} to map the the present-day kinematics of the GSE debris to their initial radii within the progenitor. Top: Cumulative radial distribution of simulated stars in the $z = 2$ GSE progenitor, selected to kinematically match observed tracers at $z = 0$: H3 GSE in gold, H3 Arjuna in purple, and our northern and southern MIKE targets in red and blue respectively. Bottom: Median observed metallicity versus median simulated GSE galactocentric radius for each  population. We fit this metallicity gradient with a linear model and display the best-fit parameters in radial and half-mass radius ($r_{50}$) coordinates.}
    \label{fig:metgrad}
\end{figure}

The distant GSE debris identified here is expected to contain the earliest-stripped stars from the outer regions of the GSE progenitor, since they retain the progenitor's large initial angular momentum prior to radialization. This correspondence between GSE disk radius and present-day angular momentum can be used to trace observed stars back to their predicted location in the simulated GSE disk. We query the $z = 0$ snapshots from \citetalias{Naidu2021} to select stars similar to our MIKE targets in Galactocentric coordinates, and trace them back to their locations in the intact $z = 2$ GSE progenitor. The resulting (cumulative) radial distributions are shown in the top panel of Figure~\ref{fig:metgrad}. We also show the corresponding distributions for simulation particles selected following the inner GSE and Arjuna selections in the H3 Survey. Clearly, these populations originate from different radial regions of the GSE progenitor. Following \citetalias{Naidu2021}, we can leverage this effect to measure the metallicity gradient of the GSE progenitor out to $3$~times its half-mass radius. 

For the GSE and Arjuna metallicities, we utilize the H3 Survey measurements from \citetalias{Naidu2021}. For the MIKE-N and MIKE-S stars, we use our \textit{Gaia} DR3 XP metallicities ($\S$\ref{sec:data}) since they were trained on H3 data and should lie on a comparable metallicity scale, whereas the CaT metallicities are systematically offset $\approx 0.1$~dex metal-poor compared to XP. {Finally, we include the all-sky retrograde debris beyond $40$~kpc described in $\S$~\ref{sec:analysis:allsky:chemistry}, again using the XP metallicities for consistency. We select correspondingly distant and retrograde particles from the \citetalias{Naidu2021} simulation, after applying the on-sky survey selection function shown in Figure~\ref{fig:lz_maps}}. 

For each sample described above, we measure the median (spherical) galactocentric radius in the simulated GSE progenitor $r_{\mathrm{GSE}}$, as well as the median observed metallicity with bootstrapped uncertainties. We display the resulting metallicity gradient in the bottom panel of Figure~\ref{fig:metgrad}, along with a linear fit that accounts for the heteroskedastic metallicity measurements. We measure a gradient of $d\mathrm{[Fe/H]}/dr = -0.02 \pm 0.01~\mathrm{dex~kpc^{-1}}$, or $d\mathrm{[Fe/H]}/dr_{50} = -0.06 \pm 0.02~\mathrm{dex~r_{50}^{-1}}$ when normalizing to the simulated GSE half-mass radius of $\approx 2.8$~kpc. 

The fit is primarily driven by the high-precision H3 datapoints, but the outer debris presented here is entirely consistent with the inferred negative metallicity gradient. Our conclusions are qualitatively unchanged if we use the MIKE CaT metallicities instead, but they are offset $\approx 0.1$~dex lower than the XP metallicities and would correspondingly lead to a slight tension with the H3-extrapolated linear trend. We emphasize that these uncertainties are purely statistical, and do not account for the (substantial) systematic uncertainty mapping these populations to $r_{\mathrm{GSE}}$ via the \citetalias{Naidu2021} N-body simulation. 

The distant debris presented here probes the very outer disk of the GSE progenitor, and it is reassuring that the metallicity gradient inferred by \citetalias{Naidu2021} extrapolates well into this regime. This is a relatively shallow gradient compared to intact dwarf galaxies around the Milky Way \citep{Kirby2011}, supporting the picture that metallicity gradients steepen over cosmic time \citep[e.g.,][]{Curti2020,Sharda2021,Tissera2022}. 
{We note that if the proposed Sequoia merger \citep{Myeong2019} is instead a metal-poor and retrograde tail of the GSE merger itself \citep[e.g.,][]{Amarante2022,Horta2022a}, then the implied metallicity gradient would be about twice as steep as that measured here \citep{Limberg2022}. 
}

\section{Discussion}\label{sec:discussion}

We have assembled a sample of luminous red giant stars out to $100$~kpc and discovered a large population of retrograde debris that we argue represent the most distant echoes of the GSE merger. 
We first established that the Outer Virgo Overdensity contains predominantly retrograde stars (Figure~\ref{fig:mike_iom}), and found a corresponding retrograde stellar population in the southern hemisphere near the previously identified Pisces Overdensity. 
These structures match predictions of early-stripped debris piling up near their apocenters, and reproduce two key aspects of the simulations --- retrograde kinematics, and a northern overdensity that is $\approx 30\%$ further than the southern one. 
We combined our data with simulations from \citetalias{Naidu2021} to map these populations back to their locations in the GSE progenitor's stellar disk, and measured a negative metallicity gradient in this $z \approx 2$ accreted galaxy out to three half-mass radii that is consistent with past measurements from the H3 Survey. 

\begin{figure*}
    \centering
    \includegraphics[width=0.4\textwidth]{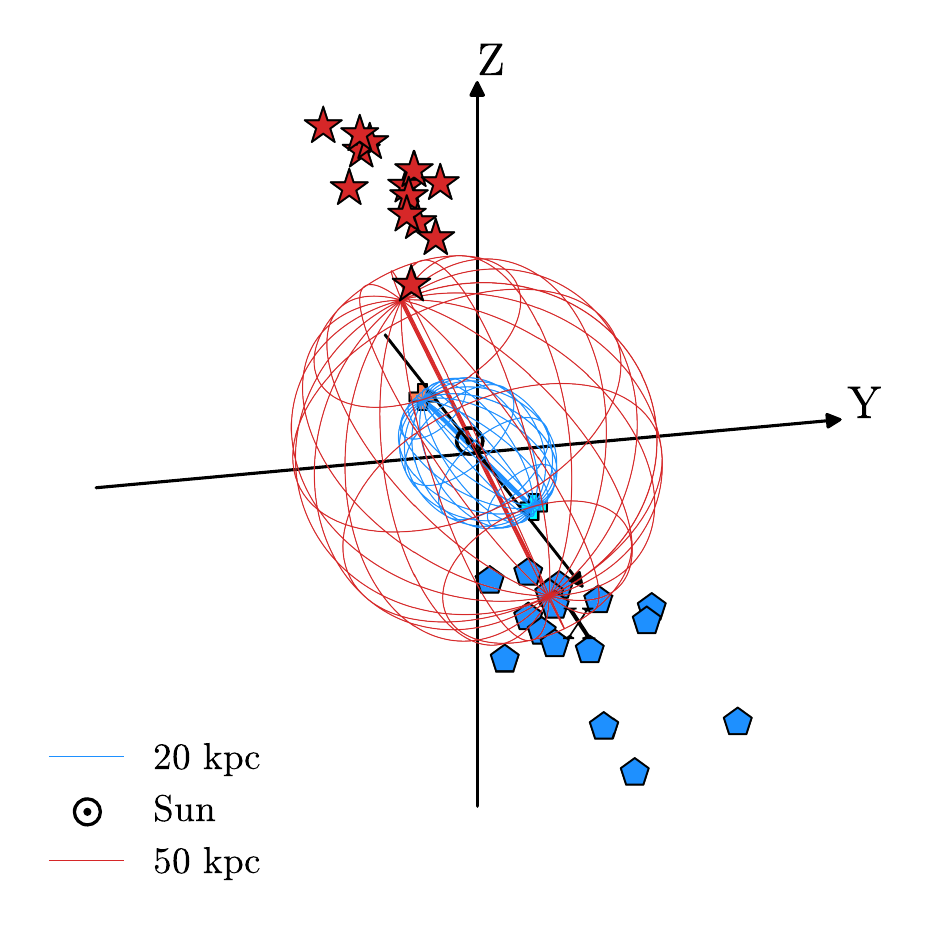}
    \includegraphics[width=0.4\textwidth]{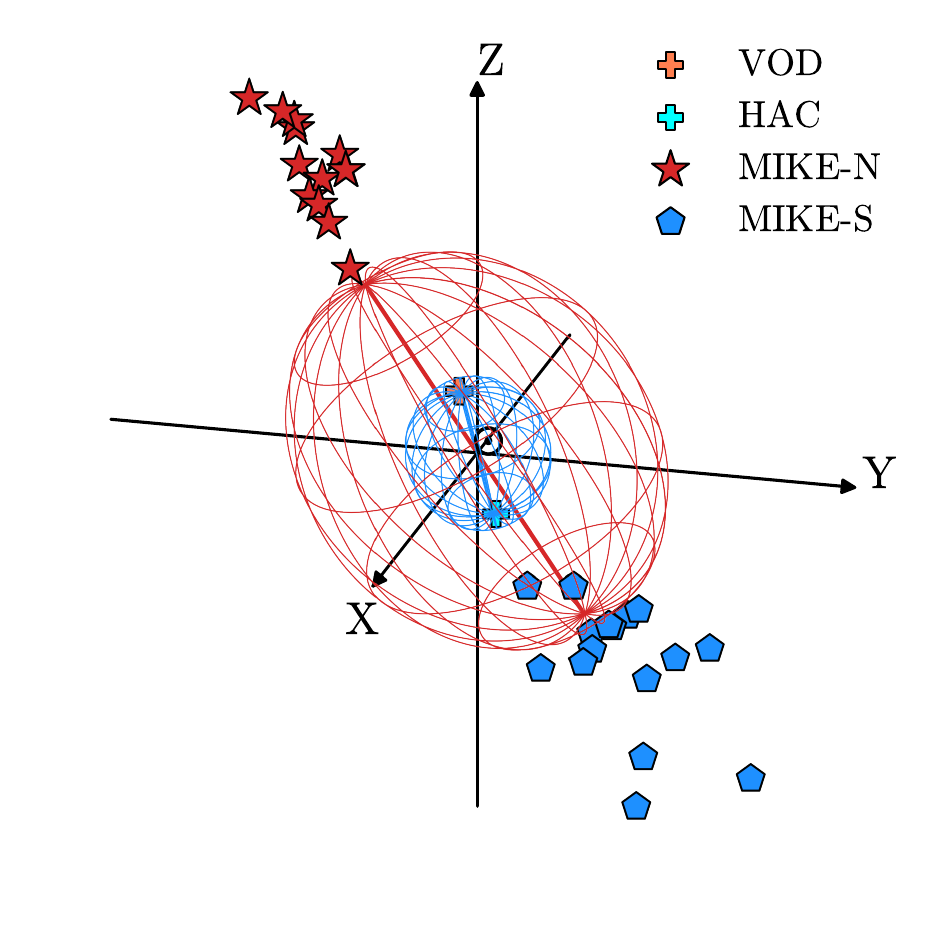}
    \includegraphics[width=\textwidth]{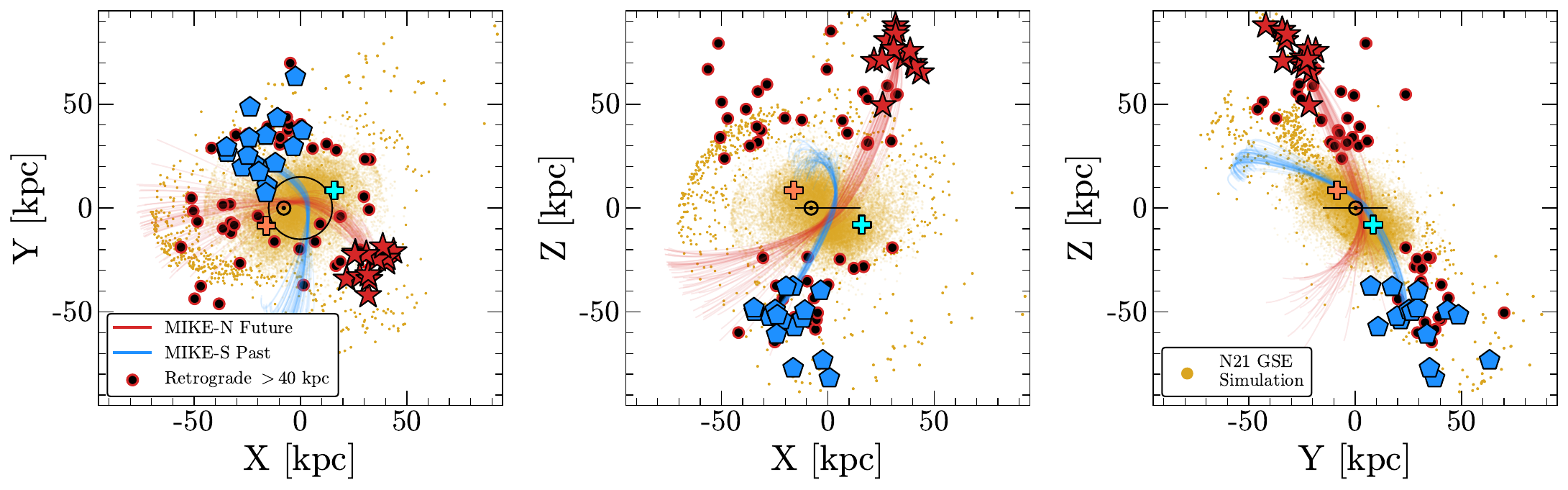}
    \caption{Galactocentric distribution of the structures described in this work. The top panels show two 3D projections of the data, whereas the bottom panels are 2D projections along each axis. The previously-identified Hercules Aquila Cloud and (inner) Virgo overdensity are shown with plus symbols using the RRL locations from \cite{Simion2019}. In the top panels we show in blue the shape of inner GSE debris fitted to H3 Survey data by \cite{Han2022b}, evaluated at $20$~kpc. Our MIKE-N sample in the Outer Virgo Overdensity and MIKE-S sample in the Pisces Overdensity are overlaid using the adopted MIKE spectroscopic distances. The red ellipsoid in the top panels is a scaled up version of the \cite{Han2022b} profile evluated at $50$~kpc, rotated to align with the MIKE-N and MIKE-S stars. In the bottom panels we also show retrograde GSE `stream` members from the 6D \textit{Gaia} XP giant sample (see Figures~\ref{fig:lz_maps} \& \ref{fig:phi1_summary}). GSE simulation particles from \citetalias{Naidu2021} at $z = 0$ are shown in gold, with particles beyond $50$~kpc emphasized. 100 realizations of the mean future (past) orbits of the MIKE-N (MIKE-S) stars are overlaid (see $\S$\ref{sec:analysis:kin} and Figure~\ref{fig:mike_iom}).}
    \label{fig:galcen_summary}
\end{figure*}

Searching for more distant retrograde stars in the broader sample of giants, we find that the majority ($\gtrsim 80\%$) of retrograde stars beyond $40$~kpc coherently lie along a stream-like track that is well-fit by a great circle on the sky (Figure~\ref{fig:lz_maps}). 
These stars can be cleanly differentiated from co-spatial Sagittarius stream stars based on their kinematics (Figures~\ref{fig:lz_maps}~\&~\ref{fig:phi1_summary}), and appear chemically consistent with GSE (Figure~\ref{fig:retro6d_tw}). 
A truly all-sky picture evades us due to the lack of spectroscopic data in the southern celestial hemisphere. 
Future spectroscopy of the distant halo in the southern hemisphere (e.g., with SDSS-V, \citealt{Kollmeier2017}) will help fill in this missing portion of the sky. 

We summarize the Galactocentric distribution of the various structures described in this work in Figure~\ref{fig:galcen_summary}, along with the median orbits of the MIKE-N and MIKE-S sample. 
The top panels show the shape of inner GSE debris in H3 Survey data \cite[][blue]{Han2022b}, along with a scaled up version rotated to match the orientation of the MIKE-N and MIKE-S stars (red). 
These panels illustrate how the $60-90$~kpc debris presented in this work relate to the previously identified GSE apocenters that form the $15-20$~kpc inner Virgo Overdensity and Hercules Aquila Cloud. 
There is a $\approx 45^\circ$ offset between the inner and outer overdensities along two axes.  
This is consistent with the \citetalias{Naidu2021} simulations, in which the inner and outer apocenters are misaligned by up to $\approx 50^\circ$ in the XY plane, due to the sharp radialization of the satellite over time. 
These four overdensities strongly constrain the in-falling trajectory of the GSE progenitor. 
A natural next step would be to self-consistently simulate the GSE merger to identify a configuration that reproduces the kinematic qualities of the inner halo, along with the spatial orientation of these four overdensities. 

\newpage

\subsection{The Emerging Picture of the GSE Merger}

In the merger simulations of \citetalias{Naidu2021}, the GSE progenitor arrived on a tilted and retrograde trajectory before rapidly radializing as it merged with the Milky Way \citep[see also][]{Bignone2019,Amarante2022,Vasiliev2022}.
The retrograde orientation of the merger was required by \citetalias{Naidu2021} to reproduce the kinematics of Arjuna, the population of metal-rich retrograde stars prominently seen in the H3 Survey \citep{Naidu2020}. 
In this work, we have discovered coherent and relatively metal-rich retrograde debris in the outer halo beyond $\gtrsim 50$~kpc, which are reasonably consistent with predictions from the \citetalias{Naidu2021} simulations. 
Our findings beyond $50$~kpc match spatial constraints of the halo's shape at closer distances --- specifically that the bulk of the halo is oriented along two preferred octants that contain the inner Virgo Overdensity and the Hercules-Aquila Cloud (Figure~\ref{fig:galcen_summary}; \citealt{Iorio2019, Han2022b}). 

The existence of a significant population of GSE debris at large $r_{\text{Gal}}$ has been a generic prediction of simulations of this merger, but the mass and kinematics of this population strongly depends on the assumed configuration. 
For example, if one invokes an initially radial merger to explain the predominantly radial inner GSE stars, then the remnant could fling a significant fraction of its mass to large distances as it merges \citep[e.g.,][]{Elias2020}. 
However, both N-body simulations and analytic calculations have shown that a merger as massive as GSE would self-radialize on short timescales, eventually creating inner debris with low angular momentum for a wide range of initial merger angular momenta \citep{Amorisco2017,Naidu2021,Vasiliev2022}. 
In more gradual merger scenarios, a smaller fraction of accreted stars would be deposited at large distances, and they would retain more angular momentum. 
It is challenging to make quantitative conclusions with our present data given the complex selection functions of the underlying surveys, as well as the use of kinematics to define the distant GSE population in the first place. 
Regardless, the presence of such a prominent population of retrograde debris in the outer halo leads us to favor the latter merger scenario, in which a retrograde GSE progenitor was gradually stripped of stars that produce the echoes we see beyond $d \gtrsim 50$~kpc. 
At closer distances, this population transitions into the phase-mixed Arjuna stars, and finally the radialized bulk of GSE debris that dominate the inner stellar halo \citep{Naidu2020,Donlon2020}. 

Our work has presented evidence for distant and early-stripped retrograde GSE debris in the outer halo beyond $40$~kpc. 
These kinematics strongly suggest that the orientation of the GSE merger was retrograde, matching results from \citetalias{Naidu2021}, and the presence of the retrograde Arjuna population in the H3 Survey from $15-30$~kpc.
Conversely, \cite{Belokurov2022b} argue for a prograde merger orientation based on the slight prograde tilt of GSE stars in their \textit{Gaia} DR3 RVS dataset, which is primarily limited within $\lesssim 15$~kpc. 
In the absence of [$\alpha$/Fe] information, the RVS sample could be contaminated by prograde stars from the in-situ halo population \citep[e.g.,][]{Bonaca2017,Bonaca2020,Belokurov2020b}.
However, a more subtle prograde tilt is also seen in the APOGEE DR17 dataset once the in-situ halo is filtered out via [$\alpha$/Fe] (V. Belokurov, private communication; \citealt{Majewski2017,Abdurro'uf2022}).
A plausible resolution to this discrepancy is that the bulk of retrograde Arjuna stars lie beyond $d \gtrsim 15$~kpc, whereas the RVS and APOGEE samples primarily contain more nearby stars. 
Therefore, the earlier-stripped retrograde component could be mostly absent from these nearer datasets, while appearing prominently in the more distant H3 Survey, and in this present work. 

Any conclusions linking present-day merger debris to the original configuration of the merger are complicated by the precession of the Milky Way disk, which would shift the angular momentum vector of the Galaxy \citep[e.g.,][]{Dodge2022,Dillamore2022}. \citetalias{Naidu2021} did not account for disk precession in their angular momentum calculations. However, they also did not model the growing of the MW disk since $z \approx 2$. It is reasonable to expect that a growing disk could suppress the degree of precession, preserving the linkage between original and present-day angular momenta. The fact that the distant debris shown here closely matches the \citetalias{Naidu2021} simulations in angular momentum --- despite neglecting the effect of precession --- could be a hint that the MW disk has only weakly precessed since $z \approx 2$. A more comprehensive test would be to self-consistently simulate the GSE merger with a growing disk, and track the evolving angular momentum of the resulting debris.

\subsection{An Outer Halo in Disequilibrium}

A broader implication of this work is apparent in our all-sky maps of the distant halo (Figure~\ref{fig:allsky}). The Milky Way's outer halo contains significant substructure, and is in fact dominated by it. 
Due to the prominence of these disequilibrium structures, any efforts to measure the Milky Way's properties in the outer halo --- from its mass to the perturbative response of its dark matter halo \citep[e.g.,][]{Garavito-Camargo2019,Garavito-Camargo2020,Shen2022} --- will require more sophisticated and time-dependent modelling \citep[e.g.,][]{Vasiliev2021a,Lilleengen2023,Koposov2022}. 

Four overdensities overwhelm our maps beyond $60$~kpc: the Sagittarius stream, the Outer Virgo Overdensity, the Pisces Overdensity, and the elongated Pisces Plume flowing from the Magellanic clouds. The former three structures are clearly visible in RRL, whereas the latter is most prominent in the RGB maps. In this work we have argued that both the Outer Virgo Overdensity and at least some part of the Pisces Overdensity can be linked to the GSE merger; the origin of the Pisces Plume remains more uncertain, particularly to what extent it comprises Magellanic debris or the dynamical friction wake of the LMC. We have shown here how precise 6D kinematics and chemistry of even a handful of members can offer key links to the origins of a structure. \\

\section{Conclusions}

{We have constructed an all-sky sample of luminous red giants out to $100$~kpc with metallicities from \textit{Gaia} DR3 XP spectra, and augmented it with radial velocities from public surveys as well as follow-up spectroscopy. The key conclusions from our work are as follows:
}
\begin{enumerate}
    \item {The outer halo beyond $\approx 40$~kpc is richly structured, with four prominent overdensities dominating the sky: the Sagittarius Stream, Outer Virgo Overdensity, Pisces Overdensity, and Pisces Plume (Figure~\ref{fig:allsky}). }

    \item {Using follow-up MIKE spectroscopy of the OVO and PO, we find both to be relatively metal-rich ([Fe/H]~$\approx -1.3$) --- similar to the bulk of the stellar halo at these distances --- and orbiting significantly retrograde relative to the MW disk. We consequently argue that both of these overdensities represent apocentric pileup of debris from the \textit{Gaia}-Sausage Enceladus (GSE) dwarf galaxy, matching key predictions from simulations of the merger (Figure~\ref{fig:mike_iom}). }

    \item {Extending our search across the sky, we find evidence for a vast stream of retrograde stars encircling the Milky Way between $40-100$~kpc, in the same plane as the Sagittarius stream albeit moving in the opposite direction (Figure~\ref{fig:lz_maps}). Together with the OVO and PO, we identify these stars as the earliest-stripped debris from the GSE merger.}

    \item {We use our sample to measure the metallicity gradient in the $z \approx 2$ GSE progenitor out to the edge of its stellar disk, exploiting the link between present-day angular momentum and progenitor galactocentric radius. Although the systematic and statistical uncertainties are large, we find a negative metallicity gradient $d\mathrm{[Fe/H]}/dr_{50} = -0.06 \pm 0.02~\mathrm{dex~r_{50}^{-1}}$, consistent with previous studies (Figure~\ref{fig:metgrad}).}
\end{enumerate}
 
{Our discovery of coherent retrograde debris beyond $40$~kpc argues for an initially retrograde orbit of the GSE progenitor, a matter of debate in the current literature. The location and kinematics of the stars presented here can be used to further constrain the orbit of the merging progenitor at early times, building towards a precise picture of the last major merger that formed the majority of the Milky Way's stellar halo. More broadly, the methods presented here to identify distant RGB stars should enable a spectroscopic census of the outer halo in 6D, finally unmasking the patchwork of accreted structure that enshrouds our Galaxy.}

\appendix

\section{MIKE Spectroscopy}

{In Table~\ref{tab:spec_mem} we list the stars with follow-up MIKE spectroscopy from the MIKE-N (Outer Virgo Overdensity) and MIKE-S (Pisces Overdensity) samples presented in this work. We provide the CaT metallicities as estimated using the calibration of \cite{Carrera2013}. }

\begin{deluxetable*}{cccccc}
\label{tab:spec_mem}
\tablecaption{Stars with follow-up spectroscopy from Magellan/MIKE}
\tablehead{\colhead{\textit{Gaia} Source ID} & \colhead{RA} & \colhead{Dec.} & \colhead{$G$} & \colhead{MIKE $v_{r,h}$} & \colhead{MIKE [Fe/H]}\\ \colhead{EDR3} & \colhead{deg} & \colhead{deg} & \colhead{mag} & \colhead{$\text{km\,s}^{-1}$} & \colhead{dex}}
\startdata
3618701644156259840 & 208.208 & -8.921 & 17.1 & -52.2 & -1.5 \\
3619451541150749696 & 210.237 & -7.164 & 16.9 & 43.5 & -1.1 \\
3629830484239963392 & 197.219 & -5.127 & 17.1 & 56.5 & -1.2 \\
3635542382722531072 & 198.599 & -5.704 & 17.1 & 58.1 & -1.3 \\
3636560530489773312 & 200.267 & -3.084 & 16.8 & 2.5 & -1.2 \\
3637038882471985536 & 201.534 & -3.407 & 17.3 & 101.3 & -1.3 \\
3637178009347886976 & 202.26 & -2.88 & 16.9 & -1.1 & -1.5 \\
3639801478451745152 & 214.924 & -7.219 & 17.1 & -38.2 & -1.8 \\
3640443314069314432 & 211.555 & -6.825 & 17.0 & -23.5 & -1.4 \\
3656976434192873088 & 208.364 & -4.257 & 16.6 & 110.6 & -1.4 \\
3659398314711482624 & 211.511 & -1.183 & 17.4 & 75.5 & -1.3 \\
3662283261424032256 & 206.258 & -0.294 & 16.6 & -8.2 & -1.3 \\
3662686709176714752 & 204.023 & -0.192 & 17.1 & 18.1 & -1.2 \\
\hline
2372535196962774912 & 13.193 & -14.292 & 16.4 & -36.8 & -1.1 \\
2431365283865625472 & 4.032 & -6.307 & 16.7 & -0.2 & -1.3 \\
2438768536173626752 & 354.67 & -8.787 & 16.9 & -55.2 & -2.4 \\
2454879993588592512 & 18.87 & -15.281 & 16.1 & 76.9 & -1.2 \\
2469715257305762560 & 17.086 & -11.468 & 16.9 & -51.3 & -1.2 \\
2529777385642667136 & 12.537 & -2.1 & 16.8 & -87.6 & -1.3 \\
2534734362017687680 & 18.063 & 0.101 & 16.3 & -118.7 & -1.3 \\
2540409830586618240 & 4.513 & -3.509 & 17.0 & -0.7 & -1.6 \\
2549636867743405952 & 12.98 & 1.578 & 16.9 & 29.3 & -1.1 \\
2552533462406653696 & 12.949 & 4.514 & 16.3 & 13.4 & -2.2 \\
2554005502317914112 & 8.786 & 4.096 & 16.7 & -70.6 & -1.8 \\
2578800279797928576 & 19.916 & 9.139 & 16.3 & -159.8 & -1.2 \\
2583539312352584960 & 18.82 & 11.716 & 15.9 & -36.6 & -2.2 \\
2657602587398571904 & 348.964 & 1.815 & 17.0 & -180.6 & -1.6 \\
2664334794016300288 & 348.748 & 6.481 & 17.2 & -121.0 & -1.2 \\
2761521550212895232 & 353.562 & 9.753 & 16.9 & -8.9 & -0.8 \\
2772570443776800384 & 1.776 & 16.206 & 16.1 & -97.3 & -1.1
\enddata
\end{deluxetable*}

\clearpage

\begin{acknowledgments}

We thank the referee for constructive feedback that significantly improved the manuscript. 
VC gratefully acknowledges a Peirce Fellowship from Harvard University. 
We thank 
Morgan Fouesneau,
Rene Andrae,
David W. Hogg,
Tom Donlon,
Guilherme Limberg,
and Will Cerny
for insightful conversations, and Vasily Belokurov for detailed feedback. 
We acknowledge Kevin Schlaufman for inspiring the `ECHOS' in the title of this work. 
We are grateful to the staff at Las Campanas Observatory --- including Yuri Beletsky, Carla Fuentes, Jorge Araya, Hugo Rivera, Alberto Past\'en, Roger Leiton, Mat\'ias D\'iaz, and Carlos Contreras --- for their invaluable assistance. 
CC and PC acknowledge support from NSF grant NSF AST-2107253. Support for this work was provided by NASA through the NASA Hubble Fellowship grant HST-HF2-51515.001-A awarded by the Space Telescope Science Institute, which is operated by the Association of Universities for Research in Astronomy, Incorporated, under NASA contract NAS5-26555. 

The computations in this paper were run on the FASRC Cannon cluster supported by the FAS Division of Science Research Computing Group at Harvard University.
This work has made use of data from the European Space Agency (ESA) mission {\it Gaia} (\url{https://www.cosmos.esa.int/gaia}), processed by the {\it Gaia} Data Processing and Analysis Consortium (DPAC, \url{https://www.cosmos.esa.int/web/gaia/dpac/consortium}). Funding for the DPAC has been provided by national institutions, in particular the institutions participating in the {\it Gaia} Multilateral Agreement. 
Funding for the Sloan Digital Sky 
Survey IV has been provided by the 
Alfred P. Sloan Foundation, the U.S. 
Department of Energy Office of 
Science, and the Participating 
Institutions. 
Guoshoujing Telescope (the Large Sky Area Multi-Object Fiber Spectroscopic Telescope LAMOST) is a National Major Scientific Project built by the Chinese Academy of Sciences. Funding for the project has been provided by the National Development and Reform Commission. LAMOST is operated and managed by the National Astronomical Observatories, Chinese Academy of Sciences.
This research has made extensive use of NASA's Astrophysics Data System Bibliographic Services.
This publication makes use of data products from the Wide-field Infrared Survey Explorer, which is a joint project of the University of California, Los Angeles, and the Jet Propulsion Laboratory/California Institute of Technology, funded by the National Aeronautics and Space Administration.

\end{acknowledgments}

\software{\texttt{numpy} \citep{Harris2020}, 
\texttt{scipy} \citep{Virtanen2020}, 
\texttt{matplotlib} \citep{Hunter2007}, 
\texttt{gala} \citep{gala,adrian_price_whelan_2020_4159870}
\texttt{MINESweeper} \citep{Cargile2020}
}

\facilities{Gaia, WISE, Magellan:Clay (MIKE), MMT (Hectochelle), Sloan, LAMOST}

\bibliography{library}

\begin{thebibliography}{}
\expandafter\ifx\csname natexlab\endcsname\relax\def\natexlab#1{#1}\fi
\providecommand{\url}[1]{\href{#1}{#1}}
\providecommand{\dodoi}[1]{doi:~\href{http://doi.org/#1}{\nolinkurl{#1}}}
\providecommand{\doeprint}[1]{\href{http://ascl.net/#1}{\nolinkurl{http://ascl.net/#1}}}
\providecommand{\doarXiv}[1]{\href{https://arxiv.org/abs/#1}{\nolinkurl{https://arxiv.org/abs/#1}}}

\bibitem[{{Abdurro'uf} {et~al.}(2022){Abdurro'uf}, {Accetta}, {Aerts}, {Silva
  Aguirre}, {Ahumada}, {Ajgaonkar}, {Filiz Ak}, {Alam}, {Allende Prieto},
  {Almeida}, \& et~al.}]{Abdurro'uf2022}
{Abdurro'uf}, {Accetta}, K., {Aerts}, C., {et~al.} 2022, ApJS, 259, 35,
  \dodoi{10.3847/1538-4365/ac4414}

\bibitem[{{Alam} {et~al.}(2015){Alam}, {Albareti}, {Allende Prieto}, {Anders},
  {Anderson}, {Anderton}, {Andrews}, {Armengaud}, {Aubourg}, {Bailey}, \&
  et~al.}]{Alam2015}
{Alam}, S., {Albareti}, F.~D., {Allende Prieto}, C., {et~al.} 2015, ApJS, 219,
  12, \dodoi{10.1088/0067-0049/219/1/12}

\bibitem[{{Amarante} {et~al.}(2022){Amarante}, {Debattista}, {Beraldo E Silva},
  {Laporte}, \& {Deg}}]{Amarante2022}
{Amarante}, J. A.~S., {Debattista}, V.~P., {Beraldo E Silva}, L., {Laporte}, C.
  F.~P., \& {Deg}, N. 2022, ApJ, 937, 12, \dodoi{10.3847/1538-4357/ac8b0d}

\bibitem[{{Amorisco}(2017)}]{Amorisco2017}
{Amorisco}, N.~C. 2017, MNRAS, 464, 2882, \dodoi{10.1093/mnras/stw2229}

\bibitem[{{Bailer-Jones}(2004)}]{Bailer-Jones2004}
{Bailer-Jones}, C.~A.~L. 2004, A\&A, 419, 385,
  \dodoi{10.1051/0004-6361:20035779}

\bibitem[{Belokurov {et~al.}(2019)Belokurov, Deason, Erkal, Koposov,
  Carballo-Bello, Smith, Jethwa, \& Navarrete}]{Belokurov2019a}
Belokurov, V., Deason, A.~J., Erkal, D., {et~al.} 2019, Monthly Notices of the
  Royal Astronomical Society: Letters, 488, L47, \dodoi{10.1093/mnrasl/slz101}

\bibitem[{Belokurov {et~al.}(2018)Belokurov, Erkal, Evans, Koposov, \&
  Deason}]{Belokurov2018a}
Belokurov, V., Erkal, D., Evans, N.~W., Koposov, S.~E., \& Deason, A.~J. 2018,
  Monthly Notices of the Royal Astronomical Society, 478, 611,
  \dodoi{10.1093/mnras/sty982}

\bibitem[{Belokurov {et~al.}(2020)Belokurov, Sanders, Fattahi, Smith, Deason,
  Evans, \& Grand}]{Belokurov2020b}
Belokurov, V., Sanders, J.~L., Fattahi, A., {et~al.} 2020, Monthly Notices of
  the Royal Astronomical Society, 494, 3880, \dodoi{10.1093/MNRAS/STAA876}

\bibitem[{{Belokurov} {et~al.}(2022){Belokurov}, {Vasiliev}, {Deason},
  {Koposov}, {Fattahi}, {Dillamore}, {Davies}, \& {Grand}}]{Belokurov2022b}
{Belokurov}, V., {Vasiliev}, E., {Deason}, A.~J., {et~al.} 2022, arXiv,
  arXiv:2208.11135.
\newblock \doarXiv{2208.11135}

\bibitem[{{Belokurov} {et~al.}(2007){Belokurov}, {Evans}, {Bell}, {Irwin},
  {Hewett}, {Koposov}, {Rockosi}, {Gilmore}, {Zucker}, {Fellhauer},
  {Wilkinson}, {Bramich}, {Vidrih}, {Rix}, {Beers}, {Schneider}, {Barentine},
  {Brewington}, {Brinkmann}, {Harvanek}, {Krzesinski}, {Long}, {Pan},
  {Snedden}, {Malanushenko}, \& {Malanushenko}}]{Belokurov2007c}
{Belokurov}, V., {Evans}, N.~W., {Bell}, E.~F., {et~al.} 2007, ApJL, 657, L89,
  \dodoi{10.1086/513144}

\bibitem[{{Bennett} \& {Bovy}(2019)}]{Bennett2019}
{Bennett}, M., \& {Bovy}, J. 2019, MNRAS, 482, 1417,
  \dodoi{10.1093/mnras/sty2813}

\bibitem[{{Bernstein} {et~al.}(2003){Bernstein}, {Shectman}, {Gunnels},
  {Mochnacki}, \& {Athey}}]{Bernstein2003}
{Bernstein}, R., {Shectman}, S.~A., {Gunnels}, S.~M., {Mochnacki}, S., \&
  {Athey}, A.~E. 2003, in Society of Photo-Optical Instrumentation Engineers
  (SPIE) Conference Series, Vol. 4841, Instrument Design and Performance for
  Optical/Infrared Ground-based Telescopes, ed. M.~{Iye} \& A.~F.~M.
  {Moorwood}, 1694--1704, \dodoi{10.1117/12.461502}

\bibitem[{Besla {et~al.}(2013)Besla, Hernquist, \& Loeb}]{Besla2013}
Besla, G., Hernquist, L., \& Loeb, A. 2013, Monthly Notices of the Royal
  Astronomical Society, 428, 2342, \dodoi{10.1093/mnras/sts192}

\bibitem[{{Bignone} {et~al.}(2019){Bignone}, {Helmi}, \&
  {Tissera}}]{Bignone2019}
{Bignone}, L.~A., {Helmi}, A., \& {Tissera}, P.~B. 2019, ApJL, 883, L5,
  \dodoi{10.3847/2041-8213/ab3e0e}

\bibitem[{{Bonaca} {et~al.}(2017){Bonaca}, {Conroy}, {Wetzel}, {Hopkins}, \&
  {Kere{\v{s}}}}]{Bonaca2017}
{Bonaca}, A., {Conroy}, C., {Wetzel}, A., {Hopkins}, P.~F., \& {Kere{\v{s}}},
  D. 2017, ApJ, 845, 101, \dodoi{10.3847/1538-4357/aa7d0c}

\bibitem[{Bonaca {et~al.}(2012)Bonaca, Geha, \& Kallivayalil}]{Bonaca2012}
Bonaca, A., Geha, M., \& Kallivayalil, N. 2012, Astrophysical Journal Letters,
  760, \dodoi{10.1088/2041-8205/760/1/L6}

\bibitem[{Bonaca {et~al.}(2020)Bonaca, Conroy, Cargile, Naidu, Johnson,
  Zaritsky, Ting, Caldwell, Han, \& van Dokkum}]{Bonaca2020}
Bonaca, A., Conroy, C., Cargile, P.~A., {et~al.} 2020, The Astrophysical
  Journal, 897, L18, \dodoi{10.3847/2041-8213/ab9caa}

\bibitem[{Brown {et~al.}(2021)Brown, Vallenari, Prusti, {De Bruijne},
  Babusiaux, Biermann, Creevey, Evans, Eyer, Hutton, Jansen, Jordi, Klioner,
  Lammers, Lindegren, Luri, Mignard, Panem, Pourbaix, Randich, Sartoretti,
  Soubiran, Walton, Arenou, Bailer-Jones, Bastian, Cropper, Drimmel, Katz,
  Lattanzi, {Van Leeuwen}, Bakker, Cacciari, Casta{\~{n}}eda, {De Angeli},
  Ducourant, Fabricius, Fouesneau, Fr{\'{e}}mat, Guerra, Guerrier, Guiraud,
  {Jean-Antoine Piccolo}, Masana, Messineo, Mowlavi, Nicolas, Nienartowicz,
  Pailler, Panuzzo, Riclet, Roux, Seabroke, Sordo, Tanga, Th{\'{e}}venin,
  Gracia-Abril, Portell, Teyssier, Altmann, Andrae, Bellas-Velidis, Benson,
  Berthier, Blomme, Brugaletta, Burgess, Busso, Carry, Cellino, Cheek,
  Clementini, Damerdji, Davidson, Delchambre, Dell'Oro,
  Fern{\'{a}}ndez-Hern{\'{a}}ndez, Galluccio, Garc{\'{i}}a-Lario,
  Garcia-Reinaldos, Gonz{\'{a}}lez-N{\'{u}}{\~{n}}ez, Gosset, Haigron,
  Halbwachs, Hambly, Harrison, Hatzidimitriou, Heiter, Hern{\'{a}}ndez,
  Hestroffer, Hodgkin, Holl, Jan{\ss}en, {Jevardat De Fombelle}, Jordan,
  Krone-Martins, Lanzafame, L{\"{o}}ffler, Lorca, Manteiga, Marchal, Marrese,
  Moitinho, Mora, Muinonen, Osborne, Pancino, Pauwels, Petit, Recio-Blanco,
  Richards, Riello, Rimoldini, Robin, Roegiers, Rybizki, Sarro, Siopis, Smith,
  Sozzetti, Ulla, Utrilla, {Van Leeuwen}, {Van Reeven}, Abbas, {Abreu
  Aramburu}, Accart, Aerts, Aguado, Ajaj, Altavilla, {\'{A}}lvarez,
  {{\'{A}}lvarez Cid-Fuentes}, Alves, Anderson, {Anglada Varela}, Antoja,
  Audard, Baines, Baker, Balaguer-N{\'{u}}{\~{n}}ez, Balbinot, Balog, Barache,
  Barbato, Barros, Barstow, Bartolom{\'{e}}, Bassilana, Bauchet,
  Baudesson-Stella, Becciani, Bellazzini, Bernet, Bertone, Bianchi,
  Blanco-Cuaresma, Boch, Bombrun, Bossini, Bouquillon, Bragaglia, Bramante,
  Breedt, Bressan, Brouillet, Bucciarelli, Burlacu, Busonero, Butkevich, Buzzi,
  Caffau, Cancelliere, C{\'{a}}novas, Cantat-Gaudin, Carballo, Carlucci,
  Carnerero, Carrasco, Casamiquela, Castellani, Castro-Ginard, {Castro Sampol},
  Chaoul, Charlot, Chemin, Chiavassa, Cioni, Comoretto, Cooper, Cornez, Cowell,
  Crifo, Crosta, Crowley, Dafonte, Dapergolas, David, David, {De Laverny}, {De
  Luise}, {De March}, {De Ridder}, {De Souza}, {De Teodoro}, {De Torres}, {Del
  Peloso}, {Del Pozo}, Delbo, Delgado, Delgado, Delisle, {Di Matteo}, Diakite,
  Diener, Distefano, Dolding, Eappachen, Edvardsson, Enke, Esquej, Fabre,
  Fabrizio, Faigler, Fedorets, Fernique, Fienga, Figueras, Fouron, Fragkoudi,
  Fraile, Franke, Gai, Garabato, Garcia-Gutierrez, Garc{\'{i}}a-Torres,
  Garofalo, Gavras, Gerlach, Geyer, Giacobbe, Gilmore, Girona, Giuffrida,
  Gomel, Gomez, Gonzalez-Santamaria, Gonz{\'{a}}lez-Vidal, Granvik,
  Guti{\'{e}}rrez-S{\'{a}}nchez, Guy, Hauser, Haywood, Helmi, Hidalgo, Hilger,
  H{\l}adczuk, Hobbs, Holland, Huckle, Jasniewicz, Jonker, {Juaristi Campillo},
  Julbe, Karbevska, Kervella, Khanna, Kochoska, Kontizas, Kordopatis, Korn,
  Kostrzewa-Rutkowska, Kruszy{\'{n}}ska, Lambert, Lanza, Lasne, {Le Campion},
  {Le Fustec}, Lebreton, Lebzelter, Leccia, Leclerc, Lecoeur-Taibi, Liao,
  Licata, Lindstr{\o}m, Lister, Livanou, Lobel, {Madrero Pardo}, Managau, Mann,
  Marchant, Marconi, {Marcos Santos}, Marinoni, Marocco, Marshall, {Martin
  Polo}, Mart{\'{i}}n-Fleitas, Masip, Massari, Mastrobuono-Battisti, Mazeh,
  McMillan, Messina, Michalik, Millar, Mints, Molina, Molinaro, Moln{\'{a}}r,
  Montegriffo, Mor, Morbidelli, Morel, Morris, Mulone, Munoz, Muraveva, Murphy,
  Musella, Noval, Ord{\'{e}}novic, Orr{\`{u}}, Osinde, Pagani, Pagano,
  Palaversa, Palicio, Panahi, Pawlak, {Pe{\~{n}}alosa Esteller},
  Penttil{\"{a}}, Piersimoni, Pineau, Plachy, Plum, Poggio, Poretti, Poujoulet,
  Pr{\v{s}}a, Pulone, Racero, Ragaini, Rainer, Raiteri, Rambaux, Ramos,
  Ramos-Lerate, {Re Fiorentin}, Regibo, Reyl{\'{e}}, Ripepi, Riva, Rixon,
  Robichon, Robin, Roelens, Rohrbasser, Romero-G{\'{o}}mez, Rowell, Royer,
  Rybicki, Sadowski, {Sagrist{\`{a}} Sell{\'{e}}s}, Sahlmann, Salgado,
  Salguero, Samaras, {Sanchez Gimenez}, Sanna, Santove{\~{n}}a, Sarasso,
  Schultheis, Sciacca, Segol, Segovia, S{\'{e}}gransan, Semeux, Shahaf,
  Siddiqui, Siebert, Siltala, Slezak, Smart, Solano, Solitro, Souami, Souchay,
  Spagna, Spoto, Steele, Steidelm{\"{u}}ller, Stephenson, S{\"{u}}veges,
  Szabados, Szegedi-Elek, Taris, Tauran, Taylor, Teixeira, Thuillot, Tonello,
  Torra, Torra, Turon, Unger, Vaillant, {Van Dillen}, Vanel, Vecchiato, Viala,
  Vicente, Voutsinas, Weiler, Wevers, Wyrzykowski, Yoldas, Yvard, Zhao, Zorec,
  Zucker, Zurbach, \& Zwitter}]{GaiaCollaboration2021}
Brown, A.~G., Vallenari, A., Prusti, T., {et~al.} 2021, Astronomy and
  Astrophysics, 649, 1, \dodoi{10.1051/0004-6361/202039657}

\bibitem[{Cargile {et~al.}(2020)Cargile, Conroy, Johnson, Ting, Bonaca, Dotter,
  \& Speagle}]{Cargile2020}
Cargile, P.~A., Conroy, C., Johnson, B.~D., {et~al.} 2020, The Astrophysical
  Journal, 900, 28, \dodoi{10.3847/1538-4357/aba43b}

\bibitem[{{Carlin} {et~al.}(2012){Carlin}, {Yam}, {Casetti-Dinescu}, {Willett},
  {Newberg}, {Majewski}, \& {Girard}}]{Carlin2012}
{Carlin}, J.~L., {Yam}, W., {Casetti-Dinescu}, D.~I., {et~al.} 2012, ApJ, 753,
  145, \dodoi{10.1088/0004-637X/753/2/145}

\bibitem[{Carrera {et~al.}(2013)Carrera, Pancino, Gallart, \& del
  Pino}]{Carrera2013}
Carrera, R., Pancino, E., Gallart, C., \& del Pino, A. 2013, Monthly Notices of
  the Royal Astronomical Society, 434, 1681, \dodoi{10.1093/mnras/stt1126}

\bibitem[{{Chambers} {et~al.}(2016){Chambers}, {Magnier}, {Metcalfe},
  {Flewelling}, {Huber}, {Waters}, {Denneau}, {Draper}, {Farrow}, {Finkbeiner},
  {Holmberg}, {Koppenhoefer}, {Price}, {Rest}, {Saglia}, {Schlafly}, {Smartt},
  {Sweeney}, {Wainscoat}, {Burgett}, {Chastel}, {Grav}, {Heasley}, {Hodapp},
  {Jedicke}, {Kaiser}, {Kudritzki}, {Luppino}, {Lupton}, {Monet}, {Morgan},
  {Onaka}, {Shiao}, {Stubbs}, {Tonry}, {White}, {Ba{\~n}ados}, {Bell},
  {Bender}, {Bernard}, {Boegner}, {Boffi}, {Botticella}, {Calamida},
  {Casertano}, {Chen}, {Chen}, {Cole}, {Deacon}, {Frenk}, {Fitzsimmons},
  {Gezari}, {Gibbs}, {Goessl}, {Goggia}, {Gourgue}, {Goldman}, {Grant},
  {Grebel}, {Hambly}, {Hasinger}, {Heavens}, {Heckman}, {Henderson}, {Henning},
  {Holman}, {Hopp}, {Ip}, {Isani}, {Jackson}, {Keyes}, {Koekemoer}, {Kotak},
  {Le}, {Liska}, {Long}, {Lucey}, {Liu}, {Martin}, {Masci}, {McLean}, {Mindel},
  {Misra}, {Morganson}, {Murphy}, {Obaika}, {Narayan}, {Nieto-Santisteban},
  {Norberg}, {Peacock}, {Pier}, {Postman}, {Primak}, {Rae}, {Rai}, {Riess},
  {Riffeser}, {Rix}, {R{\"o}ser}, {Russel}, {Rutz}, {Schilbach}, {Schultz},
  {Scolnic}, {Strolger}, {Szalay}, {Seitz}, {Small}, {Smith}, {Soderblom},
  {Taylor}, {Thomson}, {Taylor}, {Thakar}, {Thiel}, {Thilker}, {Unger},
  {Urata}, {Valenti}, {Wagner}, {Walder}, {Walter}, {Watters}, {Werner},
  {Wood-Vasey}, \& {Wyse}}]{Chambers2016}
{Chambers}, K.~C., {Magnier}, E.~A., {Metcalfe}, N., {et~al.} 2016, arXiv,
  arXiv:1612.05560.
\newblock \doarXiv{1612.05560}

\bibitem[{Choi {et~al.}(2016)Choi, Dotter, Conroy, Cantiello, Paxton, \&
  Johnson}]{Choi2016}
Choi, J., Dotter, A., Conroy, C., {et~al.} 2016, The Astrophysical Journal,
  823, 102, \dodoi{10.3847/0004-637x/823/2/102}

\bibitem[{{Clementini} {et~al.}(2022){Clementini}, {Ripepi}, {Garofalo},
  {Molinaro}, {Muraveva}, {Leccia}, {Rimoldini}, {Holl}, {Jevardat de
  Fombelle}, {Sartoretti}, {Marchal}, {Audard}, {Nienartowicz}, {Andrae},
  {Marconi}, {Szabados}, {Evans}, {Lecoeur-Taibi}, {Mowlavi}, {Musella}, \&
  {Eyer}}]{Clementini2022}
{Clementini}, G., {Ripepi}, V., {Garofalo}, A., {et~al.} 2022, arXiv,
  arXiv:2206.06278.
\newblock \doarXiv{2206.06278}

\bibitem[{{Cohen} \& {Huang}(2009)}]{Cohen2009}
{Cohen}, J.~G., \& {Huang}, W. 2009, ApJ, 701, 1053,
  \dodoi{10.1088/0004-637X/701/2/1053}

\bibitem[{{Cohen} \& {Huang}(2010)}]{Cohen2010}
---. 2010, ApJ, 719, 931, \dodoi{10.1088/0004-637X/719/1/931}

\bibitem[{Conroy {et~al.}(2018)Conroy, Bonaca, Naidu, Eisenstein, Johnson,
  Dotter, \& Finkbeiner}]{Conroy2018}
Conroy, C., Bonaca, A., Naidu, R.~P., {et~al.} 2018, The Astrophysical Journal,
  861, L16, \dodoi{10.3847/2041-8213/aacdf1}

\bibitem[{Conroy {et~al.}(2021)Conroy, Naidu, Garavito-Camargo, Besla,
  Zaritsky, Bonaca, \& Johnson}]{Conroy2021}
Conroy, C., Naidu, R.~P., Garavito-Camargo, N., {et~al.} 2021, Nature, 592,
  534, \dodoi{10.1038/s41586-021-03385-7}

\bibitem[{Conroy {et~al.}(2019{\natexlab{a}})Conroy, Naidu, Zaritsky, Bonaca,
  Cargile, Johnson, \& Caldwell}]{Conroy2019a}
Conroy, C., Naidu, R.~P., Zaritsky, D., {et~al.} 2019{\natexlab{a}}, The
  Astrophysical Journal, 887, 237, \dodoi{10.3847/1538-4357/ab5710}

\bibitem[{Conroy {et~al.}(2019{\natexlab{b}})Conroy, Bonaca, Cargile, Johnson,
  Caldwell, Naidu, Zaritsky, Fabricant, Moran, Rhee, Szentgyorgyi, Berlind,
  Calkins, Kattner, \& Ly}]{Conroy2019b}
Conroy, C., Bonaca, A., Cargile, P., {et~al.} 2019{\natexlab{b}}, The
  Astrophysical Journal, 883, 107, \dodoi{10.3847/1538-4357/ab38b8}

\bibitem[{{Cui} {et~al.}(2012){Cui}, {Zhao}, {Chu}, {Li}, {Li}, {Zhang}, {Su},
  {Yao}, {Wang}, {Xing}, {Li}, {Zhu}, {Wang}, {Gu}, {Luo}, {Xu}, {Zhang},
  {Liu}, {Zhang}, {Yang}, {Cao}, {Chen}, {Chen}, {Chen}, {Chen}, {Chu}, {Feng},
  {Gong}, {Hou}, {Hu}, {Hu}, {Hu}, {Jia}, {Jiang}, {Jiang}, {Jiang}, {Jin},
  {Li}, {Li}, {Li}, {Liu}, {Liu}, {Lu}, {Mao}, {Men}, {Qi}, {Qi}, {Shi},
  {Tang}, {Tao}, {Wang}, {Wang}, {Wang}, {Wang}, {Wang}, {Wang}, {Wang},
  {Wang}, {Wang}, {Wang}, {Wang}, {Wang}, {Xu}, {Xu}, {Yang}, {Yu}, {Yuan},
  {Yuan}, {Zhai}, {Zhang}, {Zhang}, {Zhang}, {Zhao}, {Zhou}, {Zhou}, {Zhu}, \&
  {Zou}}]{Cui2012}
{Cui}, X.-Q., {Zhao}, Y.-H., {Chu}, Y.-Q., {et~al.} 2012, RAA, 12, 1197,
  \dodoi{10.1088/1674-4527/12/9/003}

\bibitem[{{Curti} {et~al.}(2020){Curti}, {Maiolino}, {Cirasuolo}, {Mannucci},
  {Williams}, {Auger}, {Mercurio}, {Hayden-Pawson}, {Cresci}, {Marconi},
  {Belfiore}, {Cappellari}, {Cicone}, {Cullen}, {Meneghetti}, {Ota}, {Peng},
  {Pettini}, {Swinbank}, \& {Troncoso}}]{Curti2020}
{Curti}, M., {Maiolino}, R., {Cirasuolo}, M., {et~al.} 2020, MNRAS, 492, 821,
  \dodoi{10.1093/mnras/stz3379}

\bibitem[{{Das} \& {Binney}(2016)}]{Das2016}
{Das}, P., \& {Binney}, J. 2016, MNRAS, 460, 1725, \dodoi{10.1093/mnras/stw744}

\bibitem[{{Das} {et~al.}(2020){Das}, {Hawkins}, \& {Jofr{\'e}}}]{Das2020}
{Das}, P., {Hawkins}, K., \& {Jofr{\'e}}, P. 2020, MNRAS, 493, 5195,
  \dodoi{10.1093/mnras/stz3537}

\bibitem[{{De Angeli} {et~al.}(2022){De Angeli}, {Weiler}, {Montegriffo},
  {Evans}, {Riello}, {Andrae}, {Carrasco}, {Busso}, {Burgess}, {Cacciari},
  {Davidson}, {Harrison}, {Hodgkin}, {Jordi}, {Osborne}, {Pancino},
  {Altavilla}, {Barstow}, {Bailer-Jones}, {Bellazzini}, {Brown}, {Castellani},
  {Cowell}, {Delchambre}, {De Luise}, {Diener}, {Fabricius}, {Fouesneau},
  {Fremat}, {Gilmore}, {Giuffrida}, {Hambly}, {Hidalgo}, {Holland},
  {Kostrzewa-Rutkowska}, {van Leeuwen}, {Lobel}, {Marinoni}, {Miller},
  {Pagani}, {Palaversa}, {Piersimoni}, {Pulone}, {Ragaini}, {Rainer},
  {Richards}, {Rixon}, {Ruz-Mieres}, {Sanna}, {Sarro}, {Rowell}, {Sordo},
  {Walton}, \& {Yoldas}}]{DeAngeli2022}
{De Angeli}, F., {Weiler}, M., {Montegriffo}, P., {et~al.} 2022, arXiv,
  arXiv:2206.06143.
\newblock \doarXiv{2206.06143}

\bibitem[{{Deason} {et~al.}(2013){Deason}, {Belokurov}, {Evans}, \&
  {Johnston}}]{Deason2013}
{Deason}, A.~J., {Belokurov}, V., {Evans}, N.~W., \& {Johnston}, K.~V. 2013,
  ApJ, 763, 113, \dodoi{10.1088/0004-637X/763/2/113}

\bibitem[{{Dillamore} {et~al.}(2022){Dillamore}, {Belokurov}, {Font}, \&
  {McCarthy}}]{Dillamore2022}
{Dillamore}, A.~M., {Belokurov}, V., {Font}, A.~S., \& {McCarthy}, I.~G. 2022,
  MNRAS, 513, 1867, \dodoi{10.1093/mnras/stac1038}

\bibitem[{{Dodge} {et~al.}(2022){Dodge}, {Slone}, {Lisanti}, \&
  {Cohen}}]{Dodge2022}
{Dodge}, B.~C., {Slone}, O., {Lisanti}, M., \& {Cohen}, T. 2022, arXiv,
  arXiv:2207.02861.
\newblock \doarXiv{2207.02861}

\bibitem[{{Dong-P{\'a}ez} {et~al.}(2022){Dong-P{\'a}ez}, {Vasiliev}, \&
  {Evans}}]{Dong-Paez2022}
{Dong-P{\'a}ez}, C.~A., {Vasiliev}, E., \& {Evans}, N.~W. 2022, MNRAS, 510,
  230, \dodoi{10.1093/mnras/stab3361}

\bibitem[{{Donlon} {et~al.}(2020){Donlon}, {Newberg}, {Sanderson}, \&
  {Widrow}}]{Donlon2020}
{Donlon}, Thomas, I., {Newberg}, H.~J., {Sanderson}, R., \& {Widrow}, L.~M.
  2020, ApJ, 902, 119, \dodoi{10.3847/1538-4357/abb5f6}

\bibitem[{{Donlon} {et~al.}(2019){Donlon}, {Newberg}, {Weiss}, {Amy}, \&
  {Thompson}}]{Donlon2019}
{Donlon}, Thomas, I., {Newberg}, H.~J., {Weiss}, J., {Amy}, P., \& {Thompson},
  J. 2019, ApJ, 886, 76, \dodoi{10.3847/1538-4357/ab4f72}

\bibitem[{{Drimmel} \& {Poggio}(2018)}]{Drimmel2018}
{Drimmel}, R., \& {Poggio}, E. 2018, RNAAS, 2, 210,
  \dodoi{10.3847/2515-5172/aaef8b}

\bibitem[{Eisenstein {et~al.}(2011)Eisenstein, Weinberg, Agol, Aihara, {Allende
  Prieto}, Anderson, Arns, Aubourg, Bailey, Balbinot, Barkhouser, Beers,
  Berlind, Bickerton, Bizyaev, Blanton, Bochanski, Bolton, Bosman, Bovy,
  Brandt, Breslauer, Brewington, Brinkmann, Brown, Brownstein, Burger, Busca,
  Campbell, Cargile, Carithers, Carlberg, Carr, Chang, Chen, Chiappini,
  Comparat, Connolly, Cortes, Croft, Cunha, {Da Costa}, Davenport, Dawson, {De
  Lee}, {Porto De Mello}, {De Simoni}, Dean, Dhital, Ealet, Ebelke, Edmondson,
  Eiting, Escoffier, Esposito, Evans, Fan, {Femen{\'{i}}a Castell}, {Dutra
  Ferreira}, Fitzgerald, Fleming, Font-Ribera, Ford, Frinchaboy, {Garc{\'{i}}a
  P{\'{e}}rez}, Gaudi, Ge, Ghezzi, Gillespie, Gilmore, Girardi, Gott, Gould,
  Grebel, Gunn, Hamilton, Harding, Harris, Hawley, Hearty, Hennawi, {Gonzlez
  Hernndez}, Ho, Hogg, Holtzman, Honscheid, Inada, Ivans, Jiang, Jiang,
  Johnson, Jordan, Jordan, Kauffmann, Kazin, Kirkby, Klaene, Knapp, Kneib,
  Kochanek, Koesterke, Kollmeier, Kron, Lampeitl, Lang, Lawler, {Le Goff}, Lee,
  Lee, Leisenring, Lin, Liu, Long, Loomis, Lucatello, Lundgren, Lupton, Ma, Ma,
  MacDonald, MacK, Mahadevan, Maia, Majewski, Makler, Malanushenko,
  Malanushenko, Mandelbaum, Maraston, Margala, Maseman, Masters, McBride,
  McDonald, McGreer, McMahon, {Mena Requejo}, M{\'{e}}nard,
  Miralda-Escud{\'{e}}, Morrison, Mullally, Muna, Murayama, Myers, Naugle,
  Neto, Nguyen, Nichol, Nidever, O'Connell, Ogando, Olmstead, Oravetz,
  Padmanabhan, Paegert, Palanque-Delabrouille, Pan, Pandey, Parejko,
  P{\^{a}}ris, Pellegrini, Pepper, Percival, Petitjean, Pfaffenberger, Pforr,
  Phleps, Pichon, Pieri, Prada, Price-Whelan, Raddick, Ramos, Reid, Reyle,
  Rich, Richards, Rieke, Rieke, Rix, Robin, Rocha-Pinto, Rockosi, Roe,
  Rollinde, Ross, Ross, Rossetto, Snchez, Santiago, Sayres, Schiavon, Schlegel,
  Schlesinger, Schmidt, Schneider, Sellgren, Shelden, Sheldon, Shetrone, Shu,
  Silverman, Simmerer, Simmons, Sivarani, Skrutskie, Slosar, Smee, Smith,
  Snedden, Stassun, Steele, Steinmetz, Stockett, Stollberg, Strauss, Szalay,
  Tanaka, Thakar, Thomas, Tinker, Tofflemire, Tojeiro, Tremonti, {Vargas
  Mag{\~{a}}a}, Verde, Vogt, Wake, Wan, Wang, Weaver, White, White, Wilson,
  Wisniewski, Wood-Vasey, Yanny, Yasuda, Y{\`{e}}che, York, Young, Zasowski,
  Zehavi, \& Zhao}]{Eisenstein2011}
Eisenstein, D.~J., Weinberg, D.~H., Agol, E., {et~al.} 2011, Astronomical
  Journal, 142, \dodoi{10.1088/0004-6256/142/3/72}

\bibitem[{{Elias} {et~al.}(2020){Elias}, {Sales}, {Helmi}, \&
  {Hernquist}}]{Elias2020}
{Elias}, L.~M., {Sales}, L.~V., {Helmi}, A., \& {Hernquist}, L. 2020, MNRAS,
  495, 29, \dodoi{10.1093/mnras/staa1090}

\bibitem[{Erkal {et~al.}(2016)Erkal, Sanders, \& Belokurov}]{Erkal2016}
Erkal, D., Sanders, J.~L., \& Belokurov, V. 2016, Monthly Notices of the Royal
  Astronomical Society, 461, 1590, \dodoi{10.1093/mnras/stw1400}

\bibitem[{Erkal {et~al.}(2019)Erkal, Belokurov, Laporte, Koposov, Li,
  Grillmair, Kallivayalil, Price-Whelan, Evans, Hawkins, Hendel, Mateu,
  Navarro, {Del Pino}, Slater, \& Sohn}]{Erkal2019a}
Erkal, D., Belokurov, V., Laporte, C.~F., {et~al.} 2019, Monthly Notices of the
  Royal Astronomical Society, 487, 2685, \dodoi{10.1093/mnras/stz1371}

\bibitem[{{Fattahi} {et~al.}(2019){Fattahi}, {Belokurov}, {Deason}, {Frenk},
  {G{\'o}mez}, {Grand}, {Marinacci}, {Pakmor}, \& {Springel}}]{Fattahi2019}
{Fattahi}, A., {Belokurov}, V., {Deason}, A.~J., {et~al.} 2019, MNRAS, 484,
  4471, \dodoi{10.1093/mnras/stz159}

\bibitem[{{Feuillet} {et~al.}(2020){Feuillet}, {Feltzing}, {Sahlholdt}, \&
  {Casagrande}}]{Feuillet2020}
{Feuillet}, D.~K., {Feltzing}, S., {Sahlholdt}, C.~L., \& {Casagrande}, L.
  2020, MNRAS, 497, 109, \dodoi{10.1093/mnras/staa1888}

\bibitem[{{Gaia Collaboration} {et~al.}(2022{\natexlab{a}}){Gaia
  Collaboration}, {Vallenari}, {Brown}, {Prusti}, {de Bruijne}, {Arenou},
  {Babusiaux}, {Biermann}, {Creevey}, {Ducourant}, \&
  et~al.}]{GaiaCollaboration2022}
{Gaia Collaboration}, {Vallenari}, A., {Brown}, A.~G.~A., {et~al.}
  2022{\natexlab{a}}, arXiv, arXiv:2208.00211.
\newblock \doarXiv{2208.00211}

\bibitem[{{Gaia Collaboration} {et~al.}(2022{\natexlab{b}}){Gaia
  Collaboration}, {Montegriffo}, {Bellazzini}, {De Angeli}, {Andrae},
  {Barstow}, {Bossini}, {Bragaglia}, {Burgess}, {Cacciari}, {Carrasco},
  {Chornay}, {Delchambre}, {Evans}, {Fouesneau}, {Fremat}, {Garabato}, {Jordi},
  {Manteiga}, {Massari}, {Palaversa}, {Pancino}, {Riello}, {Ruz Mieres},
  {Sanna}, {Santovena}, {Sordo}, {Vallenari}, {Walton}, \&
  {DPAC}}]{Montegriffo2022b}
{Gaia Collaboration}, {Montegriffo}, P., {Bellazzini}, M., {et~al.}
  2022{\natexlab{b}}, arXiv, arXiv:2206.06215.
\newblock \doarXiv{2206.06215}

\bibitem[{Gallart {et~al.}(2019)Gallart, Bernard, Brook, Ruiz-Lara, Cassisi,
  Hill, \& Monelli}]{Gallart2019}
Gallart, C., Bernard, E.~J., Brook, C.~B., {et~al.} 2019, Nature Astronomy, 3,
  932, \dodoi{10.1038/s41550-019-0829-5}

\bibitem[{Garavito-Camargo {et~al.}(2019)Garavito-Camargo, Besla, Laporte,
  Johnston, G{\'{o}}mez, \& Watkins}]{Garavito-Camargo2019}
Garavito-Camargo, N., Besla, G., Laporte, C. F.~P., {et~al.} 2019, The
  Astrophysical Journal, 884, 51, \dodoi{10.3847/1538-4357/ab32eb}

\bibitem[{Garavito-Camargo {et~al.}(2021)Garavito-Camargo, Besla, Laporte,
  Price-Whelan, Cunningham, Johnston, Weinberg, \&
  G{\'{o}}mez}]{Garavito-Camargo2020}
---. 2021, The Astrophysical Journal, 919, 109,
  \dodoi{10.3847/1538-4357/ac0b44}

\bibitem[{{G{\'o}mez} {et~al.}(2015){G{\'o}mez}, {Besla}, {Carpintero},
  {Villalobos}, {O'Shea}, \& {Bell}}]{Gomez2015}
{G{\'o}mez}, F.~A., {Besla}, G., {Carpintero}, D.~D., {et~al.} 2015, ApJ, 802,
  128, \dodoi{10.1088/0004-637X/802/2/128}

\bibitem[{{GRAVITY Collaboration} {et~al.}(2018){GRAVITY Collaboration},
  {Abuter}, {Amorim}, {Anugu}, {Baub{\"o}ck}, {Benisty}, {Berger}, {Blind},
  {Bonnet}, {Brandner}, {Buron}, {Collin}, {Chapron}, {Cl{\'e}net}, {Coud{\'e}
  Du Foresto}, {de Zeeuw}, {Deen}, {Delplancke-Str{\"o}bele}, {Dembet},
  {Dexter}, {Duvert}, {Eckart}, {Eisenhauer}, {Finger}, {F{\"o}rster
  Schreiber}, {F{\'e}dou}, {Garcia}, {Garcia Lopez}, {Gao}, {Gendron},
  {Genzel}, {Gillessen}, {Gordo}, {Habibi}, {Haubois}, {Haug}, {Hau{\ss}mann},
  {Henning}, {Hippler}, {Horrobin}, {Hubert}, {Hubin}, {Jimenez Rosales},
  {Jochum}, {Jocou}, {Kaufer}, {Kellner}, {Kendrew}, {Kervella}, {Kok},
  {Kulas}, {Lacour}, {Lapeyr{\`e}re}, {Lazareff}, {Le Bouquin}, {L{\'e}na},
  {Lippa}, {Lenzen}, {M{\'e}rand}, {M{\"u}ler}, {Neumann}, {Ott}, {Palanca},
  {Paumard}, {Pasquini}, {Perraut}, {Perrin}, {Pfuhl}, {Plewa}, {Rabien},
  {Ram{\'\i}rez}, {Ramos}, {Rau}, {Rodr{\'\i}guez-Coira}, {Rohloff}, {Rousset},
  {Sanchez-Bermudez}, {Scheithauer}, {Sch{\"o}ller}, {Schuler}, {Spyromilio},
  {Straub}, {Straubmeier}, {Sturm}, {Tacconi}, {Tristram}, {Vincent}, {von
  Fellenberg}, {Wank}, {Waisberg}, {Widmann}, {Wieprecht}, {Wiest},
  {Wiezorrek}, {Woillez}, {Yazici}, {Ziegler}, \&
  {Zins}}]{GravityCollaboration2018}
{GRAVITY Collaboration}, {Abuter}, R., {Amorim}, A., {et~al.} 2018, A\&A, 615,
  L15, \dodoi{10.1051/0004-6361/201833718}

\bibitem[{{Han} {et~al.}(2022){Han}, {Conroy}, {Johnson}, {Speagle}, {Bonaca},
  {Chandra}, {Naidu}, {Ting}, {Woody}, \& {Zaritsky}}]{Han2022b}
{Han}, J.~J., {Conroy}, C., {Johnson}, B.~D., {et~al.} 2022, \aj, 164, 249,
  \dodoi{10.3847/1538-3881/ac97e9}

\bibitem[{Harris {et~al.}(2020)Harris, Millman, van~der Walt, Gommers,
  Virtanen, Cournapeau, Wieser, Taylor, Berg, Smith, Kern, Picus, Hoyer, van
  Kerkwijk, Brett, Haldane, del R{\'{i}}o, Wiebe, Peterson,
  G{\'{e}}rard-Marchant, Sheppard, Reddy, Weckesser, Abbasi, Gohlke, \&
  Oliphant}]{Harris2020}
Harris, C.~R., Millman, K.~J., van~der Walt, S.~J., {et~al.} 2020, Nature, 585,
  357, \dodoi{10.1038/s41586-020-2649-2}

\bibitem[{Hasselquist {et~al.}(2020)Hasselquist, Zasowski, Feuillet,
  Schultheis, Nataf, Anguiano, Beaton, Beers, Cohen, Cunha,
  Fern{\'{a}}ndez-Trincado, Garc{\'{i}}a-Hern{\'{a}}ndez, Geisler, Holtzman,
  Johnson, Lane, Majewski, Bidin, Nitschelm, Roman-Lopes, Schiavon, Smith, \&
  Sobeck}]{Hasselquist2020}
Hasselquist, S., Zasowski, G., Feuillet, D.~K., {et~al.} 2020, The
  Astrophysical Journal, 901, 109, \dodoi{10.3847/1538-4357/abaeee}

\bibitem[{{Haywood} {et~al.}(2018){Haywood}, {Di Matteo}, {Lehnert}, {Snaith},
  {Khoperskov}, \& {G{\'o}mez}}]{Haywood2018}
{Haywood}, M., {Di Matteo}, P., {Lehnert}, M.~D., {et~al.} 2018, ApJ, 863, 113,
  \dodoi{10.3847/1538-4357/aad235}

\bibitem[{Helmi {et~al.}(2018)Helmi, Babusiaux, Koppelman, Massari, Veljanoski,
  \& Brown}]{Helmi2018}
Helmi, A., Babusiaux, C., Koppelman, H.~H., {et~al.} 2018, Nature, 563, 85,
  \dodoi{10.1038/s41586-018-0625-x}

\bibitem[{{Hernquist} \& {Spergel}(1992)}]{Hernquist1992}
{Hernquist}, L., \& {Spergel}, D.~N. 1992, ApJL, 399, L117,
  \dodoi{10.1086/186621}

\bibitem[{{Horta} {et~al.}(2022){Horta}, {Schiavon}, {Mackereth}, {Weinberg},
  {Hasselquist}, {Feuillet}, {O'Connell}, {Anguiano}, {Allende-Prieto},
  {Beaton}, {Bizyaev}, {Cunha}, {Geisler}, {Garc{\'\i}a-Hern{\'a}ndez},
  {Holtzman}, {J{\"o}nsson}, {Lane}, {Majewski}, {M{\'e}sz{\'a}ros}, {Minniti},
  {Nitschelm}, {Shetrone}, {Smith}, \& {Zasowski}}]{Horta2022a}
{Horta}, D., {Schiavon}, R.~P., {Mackereth}, J.~T., {et~al.} 2022, MNRAS.tmp,
  \dodoi{10.1093/mnras/stac3179}

\bibitem[{Hunter(2007)}]{Hunter2007}
Hunter, J.~D. 2007, Computing in Science \& Engineering, 9, 90,
  \dodoi{10.1109/MCSE.2007.55}

\bibitem[{Iorio \& Belokurov(2019)}]{Iorio2019}
Iorio, G., \& Belokurov, V. 2019, Monthly Notices of the Royal Astronomical
  Society, 482, 3868, \dodoi{10.1093/mnras/sty2806}

\bibitem[{{Iorio} {et~al.}(2018){Iorio}, {Belokurov}, {Erkal}, {Koposov},
  {Nipoti}, \& {Fraternali}}]{Iorio2018}
{Iorio}, G., {Belokurov}, V., {Erkal}, D., {et~al.} 2018, MNRAS, 474, 2142,
  \dodoi{10.1093/mnras/stx2819}

\bibitem[{Johnson {et~al.}(2020)Johnson, Conroy, Naidu, Bonaca, Zaritsky, Ting,
  Cargile, Han, \& Speagle}]{Johnson2020a}
Johnson, B.~D., Conroy, C., Naidu, R.~P., {et~al.} 2020, The Astrophysical
  Journal, 900, 103, \dodoi{10.3847/1538-4357/abab08}

\bibitem[{{Johnson} {et~al.}(2022){Johnson}, {Conroy}, {Johnson}, {Peter},
  {Cargile}, {Bonaca}, {Naidu}, {Woody}, {Ting}, {Han}, \&
  {Speagle}}]{Johnson2022}
{Johnson}, J.~W., {Conroy}, C., {Johnson}, B.~D., {et~al.} 2022, arXiv,
  arXiv:2210.01816.
\newblock \doarXiv{2210.01816}

\bibitem[{Juri{\'{c}} {et~al.}(2008)Juri{\'{c}}, Ivezi{\'{c}}, Brooks, Lupton,
  Schlegel, Finkbeiner, Padmanabhan, Bond, Sesar, Rockosi, Knapp, Gunn, Sumi,
  Schneider, Barentine, Brewington, Brinkmann, Fukugita, Harvanek, Kleinman,
  Krzesinski, Long, {Neilsen, Jr.}, Nitta, Snedden, \& York}]{Juric2008}
Juri{\'{c}}, M., Ivezi{\'{c}}, {\v{Z}}., Brooks, A., {et~al.} 2008, The
  Astrophysical Journal, 673, 864, \dodoi{10.1086/523619}

\bibitem[{{Katz} {et~al.}(2022){Katz}, {Sartoretti}, {Guerrier}, {Panuzzo},
  {Seabroke}, {Th{\'e}venin}, {Cropper}, {Benson}, {Blomme}, {Haigron},
  {Marchal}, {Smith}, {Baker}, {Chemin}, {Damerdji}, {David}, {Dolding},
  {Fr{\'e}mat}, {Gosset}, {Jan{\ss}en}, {Jasniewicz}, {Lobel}, {Plum},
  {Samaras}, {Snaith}, {Soubiran}, {Vanel}, {Zwitter}, {Antoja}, {Arenou},
  {Babusiaux}, {Brouillet}, {Caffau}, {Di Matteo}, {Fabre}, {Fabricius},
  {Frakgoudi}, {Haywood}, {Huckle}, {Hottier}, {Lasne}, {Leclerc},
  {Mastrobuono-Battisti}, {Royer}, {Teyssier}, {Zorec}, {Crifo}, {Jean-Antoine
  Piccolo}, {Turon}, \& {Viala}}]{Katz2022}
{Katz}, D., {Sartoretti}, P., {Guerrier}, A., {et~al.} 2022, arXiv,
  arXiv:2206.05902.
\newblock \doarXiv{2206.05902}

\bibitem[{Kelson(2003)}]{Kelson2003}
Kelson, D. 2003, Publications of the Astronomical Society of the Pacific, 115,
  688, \dodoi{10.1086/375502}

\bibitem[{Kirby {et~al.}(2011)Kirby, Lanfranchi, Simon, Cohen, \&
  Guhathakurta}]{Kirby2011}
Kirby, E.~N., Lanfranchi, G.~A., Simon, J.~D., Cohen, J.~G., \& Guhathakurta,
  P. 2011, Astrophysical Journal, 727, \dodoi{10.1088/0004-637X/727/2/78}

\bibitem[{Kollmeier {et~al.}(2009)Kollmeier, Gould, Shectman, Thompson,
  Preston, Simon, Crane, Ivezi{\'{c}}, \& Sesar}]{Kollmeier2009}
Kollmeier, J.~A., Gould, A., Shectman, S., {et~al.} 2009, Astrophysical
  Journal, 705, \dodoi{10.1088/0004-637X/705/2/L158}

\bibitem[{Kollmeier {et~al.}(2017)Kollmeier, Zasowski, Rix, Johns, Anderson,
  Drory, Johnson, Pogge, Bird, Blanc, Brownstein, Crane, {De Lee}, Klaene,
  Kreckel, MacDonald, Merloni, Ness, O'Brien, Sanchez-Gallego, Sayres, Shen,
  Thakar, Tkachenko, Aerts, Blanton, Eisenstein, Holtzman, Maoz, Nandra,
  Rockosi, Weinberg, Bovy, Casey, Chaname, Clerc, Conroy, Eracleous,
  G{\"{a}}nsicke, Hekker, Horne, Kauffmann, McQuinn, Pellegrini, Schinnerer,
  Schlafly, Schwope, Seibert, Teske, \& van Saders}]{Kollmeier2017}
Kollmeier, J.~A., Zasowski, G., Rix, H.-W., {et~al.} 2017, in arXiv, 274.
\newblock \doarXiv{1711.03234}

\bibitem[{{Koposov} {et~al.}(2022){Koposov}, {Erkal}, {Li}, {Da Costa},
  {Cullinane}, {Ji}, {Kuehn}, {Lewis}, {Pace}, {Shipp}, {Zucker},
  {Bland-Hawthorn}, {Lilleengen}, \& {Martell}}]{Koposov2022}
{Koposov}, S.~E., {Erkal}, D., {Li}, T.~S., {et~al.} 2022, arXiv,
  arXiv:2211.04495.
\newblock \doarXiv{2211.04495}

\bibitem[{Lancaster {et~al.}(2019)Lancaster, Koposov, Belokurov, Evans, \&
  Deason}]{Lancaster2019}
Lancaster, L., Koposov, S.~E., Belokurov, V., Evans, N.~W., \& Deason, A.~J.
  2019, Monthly Notices of the Royal Astronomical Society, 486, 378,
  \dodoi{10.1093/mnras/stz853}

\bibitem[{Law \& Majewski(2010{\natexlab{a}})}]{Law2010a}
Law, D.~R., \& Majewski, S.~R. 2010{\natexlab{a}}, Astrophysical Journal, 714,
  229, \dodoi{10.1088/0004-637X/714/1/229}

\bibitem[{Law \& Majewski(2010{\natexlab{b}})}]{Law2010}
---. 2010{\natexlab{b}}, Astrophysical Journal, 718, 1128,
  \dodoi{10.1088/0004-637X/718/2/1128}

\bibitem[{{Lee} {et~al.}(2008){Lee}, {Beers}, {Sivarani}, {Allende Prieto},
  {Koesterke}, {Wilhelm}, {Re Fiorentin}, {Bailer-Jones}, {Norris}, {Rockosi},
  {Yanny}, {Newberg}, {Covey}, {Zhang}, \& {Luo}}]{Lee2008}
{Lee}, Y.~S., {Beers}, T.~C., {Sivarani}, T., {et~al.} 2008, AJ, 136, 2022,
  \dodoi{10.1088/0004-6256/136/5/2022}

\bibitem[{{Li} {et~al.}(2016){Li}, {Balbinot}, {Mondrik}, {Marshall}, {Yanny},
  {Bechtol}, {Drlica-Wagner}, {Oscar}, {Santiago}, {Simon}, {Vivas}, {Walker},
  {Wang}, {Abbott}, {Abdalla}, {Benoit-L{\'e}vy}, {Bernstein}, {Bertin},
  {Brooks}, {Burke}, {Carnero Rosell}, {Carrasco Kind}, {Carretero}, {da
  Costa}, {DePoy}, {Desai}, {Diehl}, {Doel}, {Estrada}, {Finley}, {Flaugher},
  {Frieman}, {Gruen}, {Gruendl}, {Gutierrez}, {Honscheid}, {James}, {Kuehn},
  {Kuropatkin}, {Lahav}, {Maia}, {March}, {Martini}, {Ogando}, {Plazas},
  {Reil}, {Romer}, {Roodman}, {Sanchez}, {Scarpine}, {Schubnell},
  {Sevilla-Noarbe}, {Smith}, {Soares-Santos}, {Sobreira}, {Suchyta}, {Swanson},
  {Tarle}, {Tucker}, {Zhang}, \& {DES Collaboration}}]{Li2016b}
{Li}, T.~S., {Balbinot}, E., {Mondrik}, N., {et~al.} 2016, ApJ, 817, 135,
  \dodoi{10.3847/0004-637X/817/2/135}

\bibitem[{{Lilleengen} {et~al.}(2023){Lilleengen}, {Petersen}, {Erkal},
  {Pe{\~n}arrubia}, {Koposov}, {Li}, {Cullinane}, {Ji}, {Kuehn}, {Lewis},
  {Mackey}, {Pace}, {Shipp}, {Zucker}, {Bland-Hawthorn}, {Hilmi}, \& {S5
  Collaboration}}]{Lilleengen2023}
{Lilleengen}, S., {Petersen}, M.~S., {Erkal}, D., {et~al.} 2023, MNRAS, 518,
  774, \dodoi{10.1093/mnras/stac3108}

\bibitem[{{Limberg} {et~al.}(2022){Limberg}, {Souza}, {P{\'e}rez-Villegas},
  {Rossi}, {Perottoni}, \& {Santucci}}]{Limberg2022}
{Limberg}, G., {Souza}, S.~O., {P{\'e}rez-Villegas}, A., {et~al.} 2022, ApJ,
  935, 109, \dodoi{10.3847/1538-4357/ac8159}

\bibitem[{Lindegren {et~al.}(2021)Lindegren, Klioner, Hern{\'{a}}ndez, Bombrun,
  Ramos-Lerate, Steidelm{\"{u}}ller, Bastian, Biermann, {De Torres}, Gerlach,
  Geyer, Hilger, Hobbs, Lammers, McMillan, Stephenson, Casta{\~{n}}eda,
  Davidson, Fabricius, Gracia-Abril, Portell, Rowell, Teyssier, Torra,
  Bartolom{\'{e}}, Clotet, Garralda, Gonz{\'{a}}lez-Vidal, Torra, Abbas,
  Altmann, {Anglada Varela}, Balaguer-N{\'{u}}{\~{n}}ez, Balog, Barache,
  Becciani, Bernet, Bertone, Bianchi, Bouquillon, Brown, Bucciarelli, Busonero,
  Butkevich, Buzzi, Cancelliere, Carlucci, Charlot, Cioni, Crosta, Crowley,
  {Del Peloso}, {Del Pozo}, Drimmel, Esquej, Fienga, Fraile, Gai,
  Garcia-Reinaldos, Guerra, Hambly, Hauser, Jan{\ss}en, Jordan,
  Kostrzewa-Rutkowska, Lattanzi, Liao, Licata, Lister, L{\"{o}}ffler, Marchant,
  Masip, Mignard, Mints, Molina, Mora, Morbidelli, Murphy, Pagani, Panuzzo,
  {Pe{\~{n}}alosa Esteller}, Poggio, {Re Fiorentin}, Riva, {Sagrist{\`{a}}
  Sell{\'{e}}s}, {Sanchez Gimenez}, Sarasso, Sciacca, Siddiqui, Smart, Souami,
  Spagna, Steele, Taris, Utrilla, {Van Reeven}, \& Vecchiato}]{Lindegren2021}
Lindegren, L., Klioner, S.~A., Hern{\'{a}}ndez, J., {et~al.} 2021, Astronomy
  and Astrophysics, 649, A2, \dodoi{10.1051/0004-6361/202039709}

\bibitem[{Lucchini {et~al.}(2021)Lucchini, D'Onghia, Fox, D'Onghia, \&
  Fox}]{Lucchini2021a}
Lucchini, S., D'Onghia, E., Fox, A.~J., D'Onghia, E., \& Fox, A.~J. 2021, The
  Astrophysical Journal Letters, 921, L36, \dodoi{10.3847/2041-8213/ac3338}

\bibitem[{{Mackereth} {et~al.}(2019){Mackereth}, {Schiavon}, {Pfeffer},
  {Hayes}, {Bovy}, {Anguiano}, {Allende Prieto}, {Hasselquist}, {Holtzman},
  {Johnson}, {Majewski}, {O'Connell}, {Shetrone}, {Tissera}, \&
  {Fern{\'a}ndez-Trincado}}]{Mackereth2019}
{Mackereth}, J.~T., {Schiavon}, R.~P., {Pfeffer}, J., {et~al.} 2019, MNRAS,
  482, 3426, \dodoi{10.1093/mnras/sty2955}

\bibitem[{Mainzer {et~al.}(2014)Mainzer, Bauer, Cutri, Grav, Masiero, Beck,
  Clarkson, Conrow, Dailey, Eisenhardt, Fabinsky, Fajardo-Acosta, Fowler,
  Gelino, Grillmair, Heinrichsen, Kendall, Kirkpatrick, Liu, Masci, McCallon,
  Nugent, Papin, Rice, Royer, Ryan, Sevilla, Sonnett, Stevenson, Thompson,
  Wheelock, Wiemer, Wittman, Wright, \& Yan}]{Mainzer2014}
Mainzer, A., Bauer, J., Cutri, R.~M., {et~al.} 2014, Astrophysical Journal,
  792, 30, \dodoi{10.1088/0004-637X/792/1/30}

\bibitem[{Majewski {et~al.}(2003)Majewski, Skrutskie, Weinberg, \&
  Ostheimer}]{Majewski2003}
Majewski, S.~R., Skrutskie, M.~F., Weinberg, M.~D., \& Ostheimer, J.~C. 2003,
  The Astrophysical Journal, 599, 1082, \dodoi{10.1086/379504}

\bibitem[{{Majewski} {et~al.}(2017){Majewski}, {Schiavon}, {Frinchaboy},
  {Allende Prieto}, {Barkhouser}, {Bizyaev}, {Blank}, {Brunner}, {Burton},
  {Carrera}, {Chojnowski}, {Cunha}, {Epstein}, {Fitzgerald}, {Garc{\'\i}a
  P{\'e}rez}, {Hearty}, {Henderson}, {Holtzman}, {Johnson}, {Lam}, {Lawler},
  {Maseman}, {M{\'e}sz{\'a}ros}, {Nelson}, {Nguyen}, {Nidever}, {Pinsonneault},
  {Shetrone}, {Smee}, {Smith}, {Stolberg}, {Skrutskie}, {Walker}, {Wilson},
  {Zasowski}, {Anders}, {Basu}, {Beland}, {Blanton}, {Bovy}, {Brownstein},
  {Carlberg}, {Chaplin}, {Chiappini}, {Eisenstein}, {Elsworth}, {Feuillet},
  {Fleming}, {Galbraith-Frew}, {Garc{\'\i}a}, {Garc{\'\i}a-Hern{\'a}ndez},
  {Gillespie}, {Girardi}, {Gunn}, {Hasselquist}, {Hayden}, {Hekker}, {Ivans},
  {Kinemuchi}, {Klaene}, {Mahadevan}, {Mathur}, {Mosser}, {Muna}, {Munn},
  {Nichol}, {O'Connell}, {Parejko}, {Robin}, {Rocha-Pinto}, {Schultheis},
  {Serenelli}, {Shane}, {Silva Aguirre}, {Sobeck}, {Thompson}, {Troup},
  {Weinberg}, \& {Zamora}}]{Majewski2017}
{Majewski}, S.~R., {Schiavon}, R.~P., {Frinchaboy}, P.~M., {et~al.} 2017, AJ,
  154, 94, \dodoi{10.3847/1538-3881/aa784d}

\bibitem[{{Mar{\'\i}n-Franch} {et~al.}(2012){Mar{\'\i}n-Franch}, {Chueca},
  {Moles}, {Benitez}, {Taylor}, {Cepa}, {Cenarro}, {Cristobal-Hornillos},
  {Ederoclite}, {Gruel}, {Hern{\'a}ndez-Fuertes}, {L{\'o}pez-Sainz},
  {Luis-Simoes}, {Rueda-Teruel}, {Rueda-Teruel}, {Varela}, {Yanes-D{\'\i}az},
  {Brauneck}, {Danielou}, {Dupke}, {Fern{\'a}ndez-Soto}, {Mendes de Oliveira},
  \& {Sodr{\'e}}}]{Marin-Franch2012}
{Mar{\'\i}n-Franch}, A., {Chueca}, S., {Moles}, M., {et~al.} 2012, in Society
  of Photo-Optical Instrumentation Engineers (SPIE) Conference Series, Vol.
  8450, Modern Technologies in Space- and Ground-based Telescopes and
  Instrumentation II, ed. R.~{Navarro}, C.~R. {Cunningham}, \& E.~{Prieto},
  84503S, \dodoi{10.1117/12.925430}

\bibitem[{{Montegriffo} {et~al.}(2022){Montegriffo}, {De Angeli}, {Andrae},
  {Riello}, {Pancino}, {Sanna}, {Bellazzini}, {Evans}, {Carrasco}, {Sordo},
  {Busso}, {Cacciari}, {Jordi}, {van Leeuwen}, {Vallenari}, {Altavilla},
  {Barstow}, {Brown}, {Burgess}, {Castellani}, {Cowell}, {Davidson}, {De
  Luise}, {Delchambre}, {Diener}, {Fabricius}, {Fremat}, {Fouesneau},
  {Gilmore}, {Giuffrida}, {Hambly}, {Harrison}, {Hidalgo}, {Hodgkin},
  {Holland}, {Marinoni}, {Osborne}, {Pagani}, {Palaversa}, {Piersimoni},
  {Pulone}, {Ragaini}, {Rainer}, {Richards}, {Rowell}, {Ruz-Mieres}, {Sarro},
  {Walton}, \& {Yoldas}}]{Montegriffo2022a}
{Montegriffo}, P., {De Angeli}, F., {Andrae}, R., {et~al.} 2022, arXiv,
  arXiv:2206.06205.
\newblock \doarXiv{2206.06205}

\bibitem[{Myeong {et~al.}(2019)Myeong, Vasiliev, Iorio, Evans, \&
  Belokurov}]{Myeong2019}
Myeong, G.~C., Vasiliev, E., Iorio, G., Evans, N.~W., \& Belokurov, V. 2019,
  Monthly Notices of the Royal Astronomical Society, 1247, 1235,
  \dodoi{10.1093/mnras/stz1770}

\bibitem[{{Naidu} {et~al.}(2020){Naidu}, {Conroy}, {Bonaca}, {Johnson}, {Ting},
  {Caldwell}, {Zaritsky}, \& {Cargile}}]{Naidu2020}
{Naidu}, R.~P., {Conroy}, C., {Bonaca}, A., {et~al.} 2020, ApJ, 901, 48,
  \dodoi{10.3847/1538-4357/abaef4}

\bibitem[{{Naidu} {et~al.}(2021){Naidu}, {Conroy}, {Bonaca}, {Zaritsky},
  {Weinberger}, {Ting}, {Caldwell}, {Tacchella}, {Han}, {Speagle}, \&
  {Cargile}}]{Naidu2021}
---. 2021, ApJ, 923, 92, \dodoi{10.3847/1538-4357/ac2d2d}

\bibitem[{Newberg {et~al.}(2009)Newberg, Yanny, \& Willett}]{Newberg2009}
Newberg, H.~J., Yanny, B., \& Willett, B.~A. 2009, Astrophysical Journal, 700,
  10, \dodoi{10.1088/0004-637X/700/2/L61}

\bibitem[{{Nidever} {et~al.}(2008){Nidever}, {Majewski}, \& {Butler
  Burton}}]{Nidever2008}
{Nidever}, D.~L., {Majewski}, S.~R., \& {Butler Burton}, W. 2008, ApJ, 679,
  432, \dodoi{10.1086/587042}

\bibitem[{{Nie} {et~al.}(2015){Nie}, {Smith}, {Belokurov}, {Fan}, {Fan},
  {Irwin}, {Jiang}, {Jing}, {Koposov}, {Lesser}, {Ma}, {Shen}, {Wang}, {Wu},
  {Zhang}, {Zhou}, {Zhou}, \& {Zou}}]{Nie2015}
{Nie}, J.~D., {Smith}, M.~C., {Belokurov}, V., {et~al.} 2015, ApJ, 810, 153,
  \dodoi{10.1088/0004-637X/810/2/153}

\bibitem[{Pedregosa {et~al.}(2011)Pedregosa, Varoquaux, Gramfort, Michel,
  Thirion, Grisel, Blondel, Prettenhofer, Weiss, Dubourg, Vanderplas, Passos,
  Cournapeau, Brucher, Perrot, \& Duchesnay}]{Pedregosa2011}
Pedregosa, F., Varoquaux, G., Gramfort, A., {et~al.} 2011, Journal of Machine
  Learning Research, 12, 2825

\bibitem[{{Perottoni} {et~al.}(2022){Perottoni}, {Limberg}, {Amarante},
  {Rossi}, {Queiroz}, {Santucci}, {P{\'e}rez-Villegas}, \&
  {Chiappini}}]{Perottoni2022}
{Perottoni}, H.~D., {Limberg}, G., {Amarante}, J. A.~S., {et~al.} 2022, ApJL,
  936, L2, \dodoi{10.3847/2041-8213/ac88d6}

\bibitem[{{Pop} {et~al.}(2018){Pop}, {Pillepich}, {Amorisco}, \&
  {Hernquist}}]{Pop2018}
{Pop}, A.-R., {Pillepich}, A., {Amorisco}, N.~C., \& {Hernquist}, L. 2018,
  MNRAS, 480, 1715, \dodoi{10.1093/mnras/sty1932}

\bibitem[{Price-Whelan {et~al.}(2020)Price-Whelan, Sipőcz, Lenz, Greco,
  Starkman, Foreman-Mackey, Lim, Oh, Koposov, \&
  Major}]{adrian_price_whelan_2020_4159870}
Price-Whelan, A., Sipőcz, B., Lenz, D., {et~al.} 2020, adrn/gala: v1.3, v1.3,
  Zenodo, \dodoi{10.5281/zenodo.4159870}

\bibitem[{Price-Whelan(2017)}]{gala}
Price-Whelan, A.~M. 2017, The Journal of Open Source Software, 2,
  \dodoi{10.21105/joss.00388}

\bibitem[{{Reid} \& {Brunthaler}(2004)}]{Reid2004}
{Reid}, M.~J., \& {Brunthaler}, A. 2004, ApJ, 616, 872, \dodoi{10.1086/424960}

\bibitem[{{Rix} {et~al.}(2022){Rix}, {Chandra}, {Andrae}, {Price-Whelan},
  {Weinberg}, {Conroy}, {Fouesneau}, {Hogg}, {De Angeli}, {Naidu}, {Xiang}, \&
  {Ruz-Mieres}}]{Rix2022}
{Rix}, H.-W., {Chandra}, V., {Andrae}, R., {et~al.} 2022, arXiv,
  arXiv:2209.02722.
\newblock \doarXiv{2209.02722}

\bibitem[{Ruz-Mieres(2022)}]{Mieres2022}
Ruz-Mieres, D. 2022, gaia-dpci/GaiaXPy: GaiaXPy 1.1.4, 1.1.4,  Zenodo,
  \dodoi{10.5281/zenodo.6674521}

\bibitem[{{Schlafly} \& {Finkbeiner}(2011)}]{Schlafly2011}
{Schlafly}, E.~F., \& {Finkbeiner}, D.~P. 2011, ApJ, 737, 103,
  \dodoi{10.1088/0004-637X/737/2/103}

\bibitem[{Schlafly {et~al.}(2019)Schlafly, Meisner, \& Green}]{Schlafly2019}
Schlafly, E.~F., Meisner, A.~M., \& Green, G.~M. 2019, The Astrophysical
  Journal Supplement Series, 240, 30, \dodoi{10.3847/1538-4365/aafbea}

\bibitem[{{Schlegel} {et~al.}(1998){Schlegel}, {Finkbeiner}, \&
  {Davis}}]{Schlegel1998}
{Schlegel}, D.~J., {Finkbeiner}, D.~P., \& {Davis}, M. 1998, ApJ, 500, 525,
  \dodoi{10.1086/305772}

\bibitem[{Sesar {et~al.}(2017{\natexlab{a}})Sesar, Hernitschek, Dierickx,
  Fardal, \& Rix}]{Sesar2017a}
Sesar, B., Hernitschek, N., Dierickx, M. I.~P., Fardal, M.~A., \& Rix, H.-W.
  2017{\natexlab{a}}, The Astrophysical Journal, 844, L4,
  \dodoi{10.3847/2041-8213/aa7c61}

\bibitem[{Sesar {et~al.}(2010)Sesar, Vivas, Duffau, \&
  {\v{Z}}ivezi{\'{c}}}]{Sesar2010a}
Sesar, B., Vivas, A.~K., Duffau, S., \& {\v{Z}}ivezi{\'{c}}, E. 2010,
  Astrophysical Journal, 717, 133, \dodoi{10.1088/0004-637X/717/1/133}

\bibitem[{Sesar {et~al.}(2007)Sesar, Ivezi{\'{c}}, Lupton, Juri{\'{c}}, Gunn,
  Knapp, {De Lee}, Smith, Miknaitis, Lin, Tucker, Doi, Tanaka, Fukugita,
  Holtzman, Kent, Yanny, Schlegel, Finkbeiner, Padmanabhan, Rockosi, Bond, Lee,
  Stoughton, Jester, Harris, Harding, Brinkmann, Schneider, York, Richmond, \&
  {Vanden Berk}}]{Sesar2007}
Sesar, B., Ivezi{\'{c}}, {\v{Z}}., Lupton, R.~H., {et~al.} 2007, The
  Astronomical Journal, 134, 2236, \dodoi{10.1086/521819}

\bibitem[{Sesar {et~al.}(2017{\natexlab{b}})Sesar, Hernitschek, Mitrovi{\'{c}},
  Ivezi{\'{c}}, Rix, Cohen, Bernard, Grebel, Martin, Schlafly, Burgett, Draper,
  Flewelling, Kaiser, Kudritzki, Magnier, Metcalfe, Tonry, \&
  Waters}]{Sesar2017}
Sesar, B., Hernitschek, N., Mitrovi{\'{c}}, S., {et~al.} 2017{\natexlab{b}},
  The Astronomical Journal, 153, 204, \dodoi{10.3847/1538-3881/aa661b}

\bibitem[{{Sharda} {et~al.}(2021){Sharda}, {Wisnioski}, {Krumholz}, \&
  {Federrath}}]{Sharda2021}
{Sharda}, P., {Wisnioski}, E., {Krumholz}, M.~R., \& {Federrath}, C. 2021,
  MNRAS, 506, 1295, \dodoi{10.1093/mnras/stab1836}

\bibitem[{{Shen} {et~al.}(2022){Shen}, {Eadie}, {Murray}, {Zaritsky},
  {Speagle}, {Ting}, {Conroy}, {Cargile}, {Johnson}, {Naidu}, \&
  {Han}}]{Shen2022}
{Shen}, J., {Eadie}, G.~M., {Murray}, N., {et~al.} 2022, ApJ, 925, 1,
  \dodoi{10.3847/1538-4357/ac3a7a}

\bibitem[{Simion {et~al.}(2019)Simion, Belokurov, \& Koposov}]{Simion2019}
Simion, I.~T., Belokurov, V., \& Koposov, S.~E. 2019, Monthly Notices of the
  Royal Astronomical Society, 482, 921, \dodoi{10.1093/mnras/sty2744}

\bibitem[{Simion {et~al.}(2018)Simion, Belokurov, Koposov, Sheffield, \&
  Johnston}]{Simion2018}
Simion, I.~T., Belokurov, V., Koposov, S.~E., Sheffield, A., \& Johnston, K.~V.
  2018, Monthly Notices of the Royal Astronomical Society, 476, 3913,
  \dodoi{10.1093/mnras/sty499}

\bibitem[{{Str{\"o}mgren}(1966)}]{Stromgren1966}
{Str{\"o}mgren}, B. 1966, ARA\&A, 4, 433,
  \dodoi{10.1146/annurev.aa.04.090166.002245}

\bibitem[{Ting {et~al.}(2019)Ting, Conroy, Rix, \& Cargile}]{Ting2019}
Ting, Y.-S., Conroy, C., Rix, H.-W., \& Cargile, P. 2019, The Astrophysical
  Journal, 879, 69, \dodoi{10.3847/1538-4357/ab2331}

\bibitem[{{Tissera} {et~al.}(2022){Tissera}, {Rosas-Guevara}, {Sillero},
  {Pedrosa}, {Theuns}, \& {Bignone}}]{Tissera2022}
{Tissera}, P.~B., {Rosas-Guevara}, Y., {Sillero}, E., {et~al.} 2022, MNRAS,
  511, 1667, \dodoi{10.1093/mnras/stab3644}

\bibitem[{Vasiliev {et~al.}(2021)Vasiliev, Belokurov, \& Erkal}]{Vasiliev2021a}
Vasiliev, E., Belokurov, V., \& Erkal, D. 2021, Monthly Notices of the Royal
  Astronomical Society, 501, 2279, \dodoi{10.1093/mnras/staa3673}

\bibitem[{{Vasiliev} {et~al.}(2022){Vasiliev}, {Belokurov}, \&
  {Evans}}]{Vasiliev2022}
{Vasiliev}, E., {Belokurov}, V., \& {Evans}, N.~W. 2022, ApJ, 926, 203,
  \dodoi{10.3847/1538-4357/ac4fbc}

\bibitem[{{Vincenzo} {et~al.}(2019){Vincenzo}, {Spitoni}, {Calura},
  {Matteucci}, {Silva Aguirre}, {Miglio}, \& {Cescutti}}]{Vincenzo2019}
{Vincenzo}, F., {Spitoni}, E., {Calura}, F., {et~al.} 2019, MNRAS, 487, L47,
  \dodoi{10.1093/mnrasl/slz070}

\bibitem[{Virtanen {et~al.}(2020)Virtanen, Gommers, Oliphant, Haberland, Reddy,
  Cournapeau, Burovski, Peterson, Weckesser, Bright, van~der Walt, Brett,
  Wilson, Millman, Mayorov, Nelson, Jones, Kern, Larson, Carey, Polat, Feng,
  Moore, VanderPlas, Laxalde, Perktold, Cimrman, Henriksen, Quintero, Harris,
  Archibald, Ribeiro, Pedregosa, van Mulbregt, Vijaykumar, Bardelli, Rothberg,
  Hilboll, Kloeckner, Scopatz, Lee, Rokem, Woods, Fulton, Masson,
  H{\"{a}}ggstr{\"{o}}m, Fitzgerald, Nicholson, Hagen, Pasechnik, Olivetti,
  Martin, Wieser, Silva, Lenders, Wilhelm, Young, Price, Ingold, Allen, Lee,
  Audren, Probst, Dietrich, Silterra, Webber, Slavi{\v{c}}, Nothman, Buchner,
  Kulick, Sch{\"{o}}nberger, {de Miranda Cardoso}, Reimer, Harrington,
  Rodr{\'{i}}guez, Nunez-Iglesias, Kuczynski, Tritz, Thoma, Newville,
  K{\"{u}}mmerer, Bolingbroke, Tartre, Pak, Smith, Nowaczyk, Shebanov, Pavlyk,
  Brodtkorb, Lee, McGibbon, Feldbauer, Lewis, Tygier, Sievert, Vigna, Peterson,
  More, Pudlik, Oshima, Pingel, Robitaille, Spura, Jones, Cera, Leslie, Zito,
  Krauss, Upadhyay, Halchenko, \& V{\'{a}}zquez-Baeza}]{Virtanen2020}
Virtanen, P., Gommers, R., Oliphant, T.~E., {et~al.} 2020, Nature Methods, 17,
  261, \dodoi{10.1038/s41592-019-0686-2}

\bibitem[{{Watkins} {et~al.}(2009){Watkins}, {Evans}, {Belokurov}, {Smith},
  {Hewett}, {Bramich}, {Gilmore}, {Irwin}, {Vidrih}, {Wyrzykowski}, \&
  {Zucker}}]{Watkins2009}
{Watkins}, L.~L., {Evans}, N.~W., {Belokurov}, V., {et~al.} 2009, MNRAS, 398,
  1757, \dodoi{10.1111/j.1365-2966.2009.15242.x}

\bibitem[{Xiang {et~al.}(2019)Xiang, Ting, Rix, Sandford, Buder, Lind, Liu,
  Shi, \& Zhang}]{Xiang2019}
Xiang, M., Ting, Y.-S., Rix, H.-W., {et~al.} 2019, The Astrophysical Journal
  Supplement Series, 245, 34, \dodoi{10.3847/1538-4365/ab5364}

\bibitem[{Xue {et~al.}(2015)Xue, Rix, Ma, Morrison, Bovy, Sesar, \&
  Janesh}]{Xue2015}
Xue, X.~X., Rix, H.~W., Ma, Z., {et~al.} 2015, Astrophysical Journal, 809, 144,
  \dodoi{10.1088/0004-637X/809/2/144}

\bibitem[{{Yanny} {et~al.}(2009){Yanny}, {Rockosi}, {Newberg}, {Knapp},
  {Adelman-McCarthy}, {Alcorn}, {Allam}, {Allende Prieto}, {An}, {Anderson},
  {Anderson}, {Bailer-Jones}, {Bastian}, {Beers}, {Bell}, {Belokurov},
  {Bizyaev}, {Blythe}, {Bochanski}, {Boroski}, {Brinchmann}, {Brinkmann},
  {Brewington}, {Carey}, {Cudworth}, {Evans}, {Evans}, {Gates}, {G{\"a}nsicke},
  {Gillespie}, {Gilmore}, {Nebot Gomez-Moran}, {Grebel}, {Greenwell}, {Gunn},
  {Jordan}, {Jordan}, {Harding}, {Harris}, {Hendry}, {Holder}, {Ivans},
  {Ivezi{\v{c}}}, {Jester}, {Johnson}, {Kent}, {Kleinman}, {Kniazev},
  {Krzesinski}, {Kron}, {Kuropatkin}, {Lebedeva}, {Lee}, {French Leger},
  {L{\'e}pine}, {Levine}, {Lin}, {Long}, {Loomis}, {Lupton}, {Malanushenko},
  {Malanushenko}, {Margon}, {Martinez-Delgado}, {McGehee}, {Monet}, {Morrison},
  {Munn}, {Neilsen}, {Nitta}, {Norris}, {Oravetz}, {Owen}, {Padmanabhan},
  {Pan}, {Peterson}, {Pier}, {Platson}, {Re Fiorentin}, {Richards}, {Rix},
  {Schlegel}, {Schneider}, {Schreiber}, {Schwope}, {Sibley}, {Simmons},
  {Snedden}, {Allyn Smith}, {Stark}, {Stauffer}, {Steinmetz}, {Stoughton},
  {SubbaRao}, {Szalay}, {Szkody}, {Thakar}, {Sivarani}, {Tucker}, {Uomoto},
  {Vanden Berk}, {Vidrih}, {Wadadekar}, {Watters}, {Wilhelm}, {Wyse}, {Yarger},
  \& {Zucker}}]{Yanny2009}
{Yanny}, B., {Rockosi}, C., {Newberg}, H.~J., {et~al.} 2009, AJ, 137, 4377,
  \dodoi{10.1088/0004-6256/137/5/4377}

\bibitem[{Zaritsky {et~al.}(2020)Zaritsky, Conroy, Naidu, Cargile, Putman,
  Besla, Bonaca, Caldwell, {Jesse Han}, Johnson, Speagle, \& {Ting
  丁源森}}]{Zaritsky2020}
Zaritsky, D., Conroy, C., Naidu, R.~P., {et~al.} 2020, The Astrophysical
  Journal, 905, L3, \dodoi{10.3847/2041-8213/abcb83}

\bibitem[{{Zhao} {et~al.}(2012){Zhao}, {Zhao}, {Chu}, {Jing}, \&
  {Deng}}]{Zhao2012}
{Zhao}, G., {Zhao}, Y.-H., {Chu}, Y.-Q., {Jing}, Y.-P., \& {Deng}, L.-C. 2012,
  RAA, 12, 723, \dodoi{10.1088/1674-4527/12/7/002}

\end{thebibliography}
\bibliographystyle{aasjournal}

\end{document}